\def\half{{1\over 2}}
\def\({\left (}
\def\){\right)}
\def\[{\left [}
\def\]{\right]}
\def\sqr#1#2{{\vcenter{\hrule height.#2pt\hbox{\vrule width.#2pt
height#1pt \kern#1pt \vrule width.#2pt}\hrule height.#2pt}}}
\def\square{\mathchoice\sqr64\sqr64\sqr{4.2}3\sqr{3.0}3}
\def\qed{\par\noindent\rightline{$\square$}}
\documentclass[a4paper,12pt]{article}
\usepackage[dvips]{graphicx}
\makeatletter
\renewcommand{\section}{{\setcounter{equation}{0}}\@startsection%
{section}%
{1}%
{0mm}%
{-\baselineskip}%
{0.5\baselineskip}%
{\normalfont\normalsize\bfseries}%
}
\makeatother

\newcommand{\be}{\begin{equation}}
\newcommand{\ee}{\end{equation}}
\newcommand{\bea}{\begin{eqnarray}}
\newcommand{\eea}{\end{eqnarray}}
\newcommand{\non}{\nonumber}

\font\BB=msbm10
\def\RR{\hbox{\BB R}}
\def\NN{\hbox{\BB N}}
\def\CC{\hbox{\BB C}}
\def\PP{\hbox{\BB P}}
\def\EE{\hbox{\BB E}}
\def\ZZ{\hbox{\BB Z}}
\def\QQ{\hbox{\BB Q}}
\def\Ran{{\rm Ran}}
\def\la{\langle}
\def\ra{\rangle}
\begin{document} 
\markboth{Landau Hamiltonian with Delta Impurities}
{Landau Hamiltonian with Delta Impurities}
\phantom .
\vskip3cm 
\centerline{{\bf CHARACTERIZATION OF THE SPECTRUM OF THE}}
\centerline{{\bf LANDAU HAMILTONIAN WITH DELTA IMPURITIES}}
\centerline{(To appear in Commun. Math. Phys. 1999)}
\vskip 1.5cm
\centerline{{\bf T.C. Dorlas }
\footnote{{\sl Department of Mathematics, University of Wales 
Swansea, Singleton Park, Swansea SA2 8PP, Wales, U.K.e-mail: 
T.C.Dorlas@swansea.ac.uk}}
{\bf, N. Macris }\footnote{{\sl Institut de Physique Th\'eorique,
Ecole Polytechnique F\'ed\'erale de Lausanne, CH 1015 Lausanne, Switzerland.
e-mail: macris@dpmail.epfl.ch}}
{\bf and J.V. Pul\'e }\footnote{{\sl Department of
Mathematical Physics, National University of Ireland, Dublin, 
(University College Dublin), Belfield, Dublin
4, Ireland. e-mail: Joe.Pule@ucd.ie}}
\footnote{{\sl
Research Associate, School of Theoretical 
Physics, Dublin Institute for Advanced Studies.}}
\footnote{{\sl
This work was partially supported by the Forbairt (Ireland) International Collaboration Programme 1997.}}
}
\vskip 2cm 
\begin{abstract}
\vskip -0.7truecm

\mbox{}

\noindent We consider a random 
Schr\"odinger operator in an external magnetic field. 
The random potential consists of delta functions of random strengths
situated on the sites of a regular two-dimensional lattice. 
We characterize the spectrum in the lowest $N$ Landau bands of this random Hamiltonian when the magnetic field is sufficiently strong, depending on $N$. 
We show that the spectrum in these bands is entirely pure 
point, that the energies coinciding with the Landau levels are infinitely degenerate and that the eigenfunctions corresponding to energies in the remainder of the spectrum are localized with a uniformly bounded localization length. By relating the Hamiltonian to a
lattice operator we are able to use the Aizenman-Molchanov method to prove localization. 
\end{abstract}
\newpage
\section{Introduction}
\par Recently there has been progress in the theory of Anderson localization 
for two dimensional continuous models of an electron moving in a random 
potential and a uniform magnetic field (\cite{DMP2}, \cite{DMP3}, \cite{CH}, \cite{W}). In these 
works it is established that the states at the edges of the Landau bands are 
exponentially localized and the corresponding energies form a pure point 
spectrum.  However, the nature of the generalized eigenfunctions of the 
Schr\"odinger operator for energies near the centre of the Landau bands has not 
been established. A first step towards the resolution of this problem was made in \cite{DMP1} for a Hamiltonian restricted to the first Landau band 
with a random potential consisting of point impurities with random strength 
and located on the sites of a square lattice. There it was shown that, for a sufficiently strong magnetic field, all the eigenstates are localized 
except for a single energy at the centre of the band. This energy is an 
infinitely degenerate eigenvalue with probability one.
\par In the present paper we extend the results of \cite{DMP1} to a similar model where the 
restriction to the lowest Landau band is removed. The technique used here is 
different and yields much stronger results. Formally the Hamiltonian of the 
electron is given by
\be
H = H_0 + \sum_{\bf n} v_{\bf n} \delta({\bf r - n}),
\label{a.1}
\ee
where
\be
H_0 = \half \(- i\nabla -{\bf A}({\bf r} )\)^2,
\label{a.2}
\ee
${\bf A}({\bf r})=\half ({\bf r}\times {\bf B})$ and  
$v_{\bf n}$, the strengths of the impurities which are located on the sites 
${\bf n}$ of a two-dimensional square lattice, are i.i.d. random variables. It 
is well known that the definition of Hamiltonians with point scatterer in more 
than one dimension is delicate and requires a renormalization procedure. This 
is the subject of Section 2. 
\par The main results of this paper are the following. Let $E_n = (n + \half)B$, 
$n = 0,1,2, \ldots $, be the Landau levels corresponding to the kinetic part, $H_0$, of the 
Hamiltonian. Given an integer $N$, there exists $B_0 (N)$ such that for $B> B_0 
(N)$, the spectrum is completely characterized for energies $E < E_N$. We show 
that for $n = 0, 1,2, \ldots N - 1$, the Landau levels  $E_n$ are infinitely 
degenerate eigenvalues of $H$ with probability one. All other energies in this part of the spectrum correspond to exponentially localized eigenfunctions 
with a localization length which is uniformly bounded as a function of the 
energy. Thus the localization length does not diverge at the centres of the bands when the magnetic field is strong enough, at least for the lower bands. Our analysis breaks down for energies greater than $E_N$ and in fact we expect a 
different behaviour for high energies. 
\par
There is an extensive literature on the problem of point scatterers with a 
magnetic field, but it appears that little is known on the rigorous level
for the two-dimensional random case considered here. For the periodic case,
that is, when all the $v_{\bf n}$'s are identical, we refer the reader 
to the review \cite{G} and the references therein. The case when the 
potential is periodic in the $x$-direction and random in the $y$-direction
has been discussed recently in \cite{GZAA}. Finally the density of states 
for models similar to ours with a restriction to the first Landau level
has been computed analytically in \cite{BGI} (see also \cite{E} which deals 
with the existence of Lifshitz tails). 
The infinite degeneracy of the Landau levels had already been noticed 
in various ways in the past (\cite{H}, \cite{BGI}, \cite{BF}). For example
in \cite{BGI} it appears as a delta function in the density of states of 
the first level. The result suggests that it is in fact {\sl macroscopic},
in other words, there is a positive density per unit volume. Our results
characterize completely the rest of the spectrum and also give information 
about the localization length.
\par 
Let us say a few words about the method used to arrive at these results. 
The scatterers in (\ref{a.1}) are similar to rank one perturbations of the kinetic energy so that by using the resolvent identity one can express the 
Green's function corresponding to $H$ in terms of the Green's function of the kinetic energy and a matrix which contains all the randomness. Thus the problem is reduced completely to the study of this random matrix which has random elements on the diagonal and rapidly decaying non-random off-diagonal elements. It turns out that the method invented by Aizenman and Molchanov \cite{AM} is very well suited to study the decay of eigenvectors of this matrix. 
These eigenvectors are related by an explicit formula to the eigenfunctions of $H$ in such a way that exponential decay of the former implies exponential decay of the latter. In fact it follows from the structure of the random matrix that, in the strong magnetic field regime, the off-diagonal elements are much smaller than the diagonal elements, and this is true even for energies near the band centres. Therefore our problem is 
analagous to the high disorder regime in the usual Anderson model and this is the reason why we have access to the whole spectrum.
\par It is instructive to discuss the physical implications of our results in 
the context of the quantum Hall effect. A basic ingredient used to explain the 
occurrence of plateaux in the Hall conductivity is the localization of 
electrons due to the random potential. This has been established in a 
mathematically precise way in \cite{Ku} (see also \cite{Be}), by assuming 
the existence of localized states. Usually it is difficult to obtain 
quantitative results on the localization length. The Network Model of Chalker 
and Coddington \cite{CC} and numerical simulations \cite{H} suggest that it is 
finite except at the band centres where it diverges like $\vert E - E_n 
\vert^{ - \nu}$ with $\nu \approx 2\cdot 35$ for the first few $n$'s. In the 
Network Model one must work with smooth equipotential lines of the random 
potential so that it is difficult to compare to our situation. The model in this paper has been treated numerically only in a regime where the magnetic length, which is of the order of $B^{-1/2}$, is much greater than the average spacing between impurities. The regime 
covered by our analysis is such that the magnetic length is smaller than the 
average spacing between impurities, and we prove that there is no divergence in the 
localization length at least for the first few bands. One might think that this means that
there is no quantum Hall effect in this regime. However this is not the case 
because the energy at the band centre is an infinitely degenerate eigenvalue. 
One can compute explicitly the eigenprojector associated to each degenerate
eigenvalue and check that the corresponding Chern number
is equal to unity \cite{DMP4}. From this result and the equivalence between Hall conductivity and 
Chern number, when the Fermi level lies in the region of localized 
states or in a spectral gap, we conclude that the Hall conductivity takes 
a non-zero quantized value equal to the number of Landau levels below the 
Fermi energy. This has made mathematically precise
in \cite{Ku} (see also \cite{Be} and \cite{AG}). 
\par The picture which emerges out of the combination of our analytical results 
with those of simulations is that in the present model one has to distinguish 
at least two regimes. In the first one, the magnetic length is much greater 
than the spacing between impurities: the localization length diverges and 
there is no degenerate eigenvalue at the band centres. In the second the 
magnetic length is much smaller than the spacing between impurities: the 
localization length does not diverge and there is a degenerate eigenvalue at 
the band centres. Whether there exists one or more intermediate regimes or not is an 
open question. It is instructive to note that in the model studied in \cite{BGI},
it turns out that, at the level of the density of states, one must also distinguish between various regimes, more than two in fact. 
Finally, we wish 
to stress that the quantized Hall plateaux exist in both regimes and that an 
interesting open question is whether the different behaviour of 
the localization length is reflected in the transition between  two successive 
Hall plateaux.
\par The paper is organized as follows. In Section 2 we give the precise 
definition of the model and the Hamiltonian and also collect useful Green's function 
identities. Our main theorem (Theorem 2.2) is stated at the end of this 
section. The infinite degeneracy of the first $N(B)$ Landau levels is proved 
in Section 3 and the spectrum is characterized as a set. The connection 
between generalized eigenfunctions of $H$ and eigenvectors of the random 
matrix is established in Section 4. Finally, the 
Aizenman-Molchanov method is applied in Section 5 where the proof of our 
main theorem is completed. The appendices contain more technical material.
%
%
\section{Definition of the Hamiltonian}
In this section we define our Hamiltonian. It is well known that Hamiltonians
with $\delta$-function potentials in dimensions greater than one require renormalization. This was first done rigorously in \cite{BF}. The magnetic field case was developed in \cite{G}. We refer the reader also to \cite{AGHKH} though this does not deal explicitly with the case of a magnetic field. 
\par
Let $\omega_n$, $n \in \ZZ[i]\equiv \{n_1+in_2: (n_1,n_2)\in \ZZ^2\}$, the 
Gaussian integers,
be i.i.d. random variables. We shall assume that their distribution is
given by an absolutely continuous 
probability measure $\mu_0$ whose support is an interval 
$X=[-a,a]$ with $0<a <\infty $. We require that $\mu_0$ is symmetric about the origin and that its density $\rho_0$ is differentiable on $(-a,a)$ and 
satisfies the following condition
\be
\sup_{\zeta\in (0,a)} \frac{\rho_0'(\zeta)} {\rho_0(\zeta)}< \infty.
\ee
These conditions on $\mu_0$ can be weakened, but we have chosen the above because they allow us to check the regularity of the distribution of 
$1/\omega_n $, in the sense of \cite{AM} very simply.
We let $\Omega=X^{\ZZ[i]}$ and $\PP=\prod_{n\in \ZZ[i]}\mu_0$.
For $m\in \ZZ[i]$ let $\tau_m$ be the measure preserving automorphism of 
$\Omega$ defined by 
\be\label{b.1} 
\(\tau_m \omega\)_n =\omega_{n-m}.
\ee                
The group $\{\tau_m\ :\ m\in\ZZ[i]\}$ is ergodic for the probability measure
$\PP$.
\par 
Let ${\cal H}=L^2(\CC)$ and let $H_0$ be the operator on ${\cal H}$
defined by 
\be\label{b.2}
H_0=(1/8\kappa)(-i\nabla-A(z))^2 -1/2
\ee
where $A(z)=(-2\kappa {\cal I}z, 2\kappa {\cal R}z)$.
Here $\kappa=B/4$ and $H_0$ is the same as the Hamiltonian in (\ref{a.2})
apart from the multiplicative constant $1/8\kappa$ and the shift by 
$1/2$ which are inserted for convenience so 
that the Landau levels coincide with the set of non-negative integers, 
$\NN_0$.
Let ${\cal H}_m$ be the eigenspace corresponding to the $m$th Landau 
level of the Hamiltonian $H_0$ defined in (\ref{b.2})
and let $P_m$ be the orthogonal projection onto ${\cal H}_m$. 
The projection $P_m$ is an integral operator with kernel
\be\label{b.4}
P_m \(z,z'\) = L_m(2\kappa | z-z'|^2)P_0 \(z,z'\),
\ee    
where $L_m$ is the Laguerre polynomial of order $m$ and
\be\label{b.5}
P_0 \(z,z'\) = {{2\kappa} \over { \pi}} \exp [- {\kappa} | z - z' |^2 - 
2i \kappa z \wedge z'],
\ee                      
with 
$ z  \wedge z'={\cal R}z{\cal I}z'-{\cal I}z{\cal R}z'$, 
${\cal R}z$ and ${\cal I}z$ being the real and imaginary parts of $z$
respectively.
\par 
For $\lambda \in \CC\setminus \NN_0$, let $G^\lambda_0=(H_0-\lambda)^{-1}$,
the resolvent of $H_0$ at $\lambda$.  $G^\lambda_0$ has kernel (cf \cite{G})
\be
G^\lambda_0 (z,z') 
=  \Gamma (-\lambda) P_0 (z,z') 
U (- \lambda, 1, 2 \kappa | z  - z' |^2 ),
\label{b.0}
\ee
where
\be
U(a, 1, \rho)= - {1\over {\Gamma (a) }}\[M(a, 1, \rho) \ln \rho 
+ \sum^\infty_{r=0} 
{{(a)_r} \over {r!}} \rho^r \{\psi(a + r) - 2 \psi (1 + r)\}\] 
\ee
is the logarithmic solution of Kummer's equation (\cite{AS} Chap. 13):
\be
\rho{{d^2U}\over{d\rho^2}}+(1-\rho) {{dU}\over{d\rho}}-a\rho=0
\ee
Here $\Gamma$ is the Gamma function, $\psi(a)=\Gamma'(a)/\Gamma(a)$ is the Digamma function,
\be
(a)_r=a(a+1)(a+2)\ldots (a+r-1),\ \ \ \ \ \ \ \ (a)_0=1,
\ee
and 
\be
M(a, 1, \rho)= \sum^\infty_{r=0} {{(a)_r} \over {r!}} \rho^r 
\ee
is Kummer's function.
\par
Let ${\cal M}=l^2 ([\ZZ [i]) $ and for $\lambda \in \CC\setminus \NN_0$,
define $U_\lambda:{\cal H}\to {\cal M}$
by 
\be 
\la n | U_\lambda \phi\ra = (G^\lambda_0 \phi)(n).
\ee
From the bounds in Propositions 6.1 and 6.2 in Appendix A one can see that $U_\lambda$ is a bounded operator. Its adjoint 
$U_\lambda^*:{\cal M}\to {\cal H}$ is given by
\be 
(U_\lambda^* \xi)(z) = \sum_{n\in \ZZ[i]}G^{\bar \lambda}_0(z,n)
\la n | \xi\ra.
\ee
For $\lambda \in \CC \setminus \NN_0$ let
\be
c^\lambda_n = {{2 \kappa}\over\pi} 
\(\psi (- \lambda) -{{2 \pi} \over {\omega_n}}\)
\ee
and define the operators $D^\lambda$, $A^\lambda$ and $M^\lambda$ 
on ${\cal M}$ as follows. $D^\lambda$ is diagonal and
\be
\la n | D^\lambda | n \ra =  c^\lambda_n, 
\ee
\be
\la n | A^\lambda | n' \ra 
= \cases{ 0 & if $ n = n'$ \cr
G^\lambda_0 (n,n') & if $ n \not= n'$, \cr}
\ee
and
\be
M^\lambda = D^\lambda -A^\lambda.
\label{b.3}
\ee
Note that $D^\lambda$ is a closed operator on the domain
\be
D(D^\lambda)=\{\xi\in {\cal M}:\ \ \sum_{n\in\ZZ[i]}| c^\lambda_n |^2
|\la n| \xi\ra|^2 <\infty\},
\ee
and $A^\lambda$ is bounded, therefore $M^\lambda$ is closed
on $ D(M^\lambda)=D(D^\lambda)$. Note also that 
$(M^\lambda)^*=M^{\bar \lambda}$ and that for $\lambda \in \RR$, 
$M^\lambda$ is self-adjoint.
For $\lambda \in \CC\setminus \NN_0$ such that $0\notin \sigma(M^\lambda)$ 
let 
\be
\Gamma^\lambda = (M^\lambda)^{-1}.
\ee  
To define our Hamiltonian $H$ we use the following lemma. :
\par\noindent
{\bf Lemma 2.1}\ {\sl For each $\kappa>0$, there exists $\lambda_\kappa\in 
\CC\setminus \RR$ such that   
$0\notin 
\sigma(M^{\lambda_\kappa})$ 
and 
\be
| \la n | \Gamma^{\lambda_\kappa} | n' \ra | 
\leq K(\kappa) e^{-\kappa | n - n' |^\half}.
\ee
}  
\par\noindent
{\bf Proof:}\ 
Let $\lambda=-r(1+i)$, with $r>0$. By Proposition 6.1 in Appendix A, we have for $n, \ n' \in \ZZ[i]$, $n\neq n'$,
\be
| G_0^\lambda (n,n') | 
\leq C_{r,\kappa} e^{- \kappa | n - n' |^2},
\ee
where 
\be
C_{r,\kappa}=C\kappa\left \{ {1\over r}+ e^{-(2\kappa r)^\half }(1+|\ln (2\kappa)|)\right \},
\ee
$C<\infty$ being a constant.
Therefore  $|| A^\lambda|| \leq C_{r,\kappa}|| S||$ 
where $S$ is the operator with matrix
\be
\la n| S| n'\ra = e^{-\kappa | n - n' |^2}.
\ee
Let ${\tilde \Gamma}^\lambda=(D^\lambda)^{-1}$, then
\be
|| A^\lambda{\tilde \Gamma}^\lambda|| \leq {\pi\over{2\kappa}}
{{ C_{r,\kappa}}\over {| {\cal I}\psi(-\lambda)|}}
|| S|| <1/2,
\ee
if $r$ is large enough. Note that by (6.3.18) in \cite{AS}
\be
\lim_{r\to\infty}{\cal I}\psi(-\lambda)=\pi/4. 
\ee
Then $ \sum^\infty_{k=1} (A^\lambda {\tilde \Gamma}^\lambda)^k$
converges and consequently $M^\lambda$ is invertible, 
\be\label{b.6}
\Gamma^\lambda 
= {\tilde \Gamma}^\lambda (I + \sum^\infty_{k=1} 
(A^\lambda {\tilde \Gamma}^\lambda)^k) 
\ee
and 
$|| I + \sum^\infty_{k=1} (A^\lambda {\tilde \Gamma}^\lambda)^k ||
\leq 2$. Clearly
\be
\la n | A^\lambda {\tilde \Gamma}^\lambda | n' \ra  
= \cases{0 & for $n = n'$ \cr
{1 \over {c_{n'}^\lambda}} \ \ G_0^\lambda (n,n') & if $n \not= n'$ .\cr}
\ee
Thus
\be
| \la n | A^\lambda {\tilde \Gamma}^\lambda | n' \ra | 
\leq B_{r,\kappa}e^{- \kappa | n - n' |^2}
\leq B_{r,\kappa}e^{- \kappa | n - n' |^\half}
\label{b.11}
\ee
where $ B_{r,\kappa} = {\pi\over{2\kappa}}
{{ C_{r,\kappa} } \over {| {\cal I}\psi(- \lambda)|}} $.
Now, there exists a constant $ c_0<\infty $ such that for $\kappa >1$ 
(see Lemma 3.3 in \cite{DMP1}),
\be
\sum_{n'' \in \ZZ[i] } e^{- \kappa | n - n'' |^\half}
e^{- \kappa | n'' - n' |^\half}\leq
c_0 e^{- \kappa | n - n' |^\half}.
\ee
This bound, together with (\ref{b.11}), gives 
\be
| \la n | (A^\lambda {\tilde \Gamma}^\lambda)^k | n'\ra |
\leq c_0^{k-1} B_{r,\kappa}^k 
e^{- \kappa | n - n' |^\half}
\ee
and thus from (\ref{b.6})
\be
| \la n | \Gamma^\lambda | n' \ra | 
\leq K e^{-\kappa | n - n' |^\half}
\ee
if $c_0 B_{r,\kappa} < \half$. 
\qed
\par
For $\lambda \in \CC\setminus\NN_0$, we have the formula (\cite{AS} 6.3.16) 
\be
\psi(- \lambda) = - \gamma - \sum^\infty_{m=0} {{(\lambda +1)} 
\over {(m+1)(m- \lambda)}}
\ee
where $\gamma$ is Euler's constant.
Thus if $\lambda_1, \lambda_2 \in \CC\setminus\NN_0 $ 
\be
\psi(- \lambda_1) - \psi(- \lambda_2)  = (\lambda_2 - \lambda_1) 
\sum^\infty_{m=0} {1 \over {(m - \lambda_1)(m - \lambda_2)}}.
\ee
On the other hand we have 
\be
G^{\lambda_1}_0 G^{\lambda_2}_0 = \sum^\infty_{m=0} {{P_m} \over {(m - 
\lambda_1) (m - \lambda_2)}}
\ee
and thus 
\be
(G^{\lambda_1}_0 G^{\lambda_2}_0) (n, n) = \sum^\infty_{m=0} {{P_m (n,n)} 
\over {(m - \lambda_1)(m - \lambda_2)}}
= {{2 \kappa} \over \pi} \sum^\infty_{m=0} {1 \over {(m - \lambda_1)(m - 
\lambda_2)}}.
\label{b.10}
\ee
Therefore
\be
\la n | M^{\lambda_1} - M^{\lambda_2} | n \ra = {{2 \kappa} \over 
\pi} \{\psi(- \lambda_1) - \psi(-\lambda_2)\}
= (\lambda_2 - \lambda_1)(G_0^{\lambda_1} G_0^{\lambda_2})(n,n).
\label{b.7}
\ee
On the other hand, for $n \not= n'$, using the resolvent identity, we get 
\be\label{b.8}
\la n | M^{\lambda_1} - M^{\lambda_2} | n' \ra  = 
G_0^{\lambda_2} (n,n') - G^{\lambda_1}_0 (n,n') 
= (\lambda_2 - \lambda_1)(G_0^{\lambda_1} G_0^{\lambda_2})(n,n'). 
\ee
Therefore combining the two identities (\ref{b.7}) and (\ref{b.8}) we obtain
\be\label{b.9}
M^{\lambda_1} - M^{\lambda_2} = (\lambda_2 - \lambda_1) 
U_{\lambda_2}U_{{\bar\lambda}_1}^*.
\ee
It is clear from this equation that $U_{\lambda_2}U_{\bar{\lambda}_1}^* =
U_{\lambda_1}U_{{\bar\lambda}_2}^*$.
\par
Note that $H_0$ is essentially self-adjoint on ${\cal S}(\CC)$ 
(\cite{RS2} Theorem X.34). 
Define $V_\kappa:{\cal S}(\CC)\to {\cal H}$ 
by $V_\kappa=U^*_{{{\bar \lambda}_\kappa}}\Gamma^{\lambda_\kappa}T$ where 
$\la n | T\psi \ra=\psi(n)$. Let
\be
D(H)=\{\phi=\psi +V_\kappa\psi\ : \ \psi \in {\cal S}(\CC)\},
\ee
and for $\phi \in D(H)$
\be
H\phi=H_0\psi +{\lambda_\kappa}V_\kappa\psi.
\ee
This definition implies that 
$(H-{\lambda_\kappa})\phi=(H_0-{\lambda_\kappa})\psi$ and 
therefore since $H_0$ is essentially
self-adjoint on ${\cal S}(\CC)$, $\Ran (H-{\lambda_\kappa})$ is dense 
in ${\cal H}$.
Let $\psi'\in {\cal S}(\CC)$ and let 
$\psi=\psi'+({\bar\lambda}_\kappa-\lambda_\kappa)
G_0^{\lambda_\kappa}U^*_{\lambda_\kappa}
\Gamma^{{{\bar \lambda}_\kappa}}T\psi'$. Then $\psi \in {\cal S}(\CC)$ and
$T\psi=M^{\lambda_\kappa}\Gamma^{{{\bar \lambda}_\kappa}}T\psi'$.
Note that $0\notin \sigma(M^{{\bar \lambda}_\kappa})$ 
and 
$ | \la n | \Gamma^{{\bar \lambda}_\kappa} | n' \ra |
=| \la n' | \Gamma^{\lambda_\kappa} | n \ra |
\leq K(\kappa) e^{-\kappa | n - n' |^\half}
$.
Let $\phi=\psi+V_\kappa\psi$. Then
\bea
(H-{{\bar \lambda}_\kappa})\phi 
&=& (H_0-{{\bar \lambda}_\kappa})\psi 
+(\lambda_\kappa-{\bar\lambda}_\kappa) V_\kappa\psi\non \\
&=& (H_0-{{\bar \lambda}_\kappa})\psi' 
-(\lambda_\kappa-{\bar\lambda}_\kappa) 
G_0^{\lambda_\kappa}(H_0-{{\bar \lambda}_\kappa})U^*_{\lambda_\kappa}
\Gamma^{{{\bar \lambda}_\kappa}}T\psi'\non \\
& &\ \ \ \ \ \ +(\lambda_\kappa-{\bar\lambda}_\kappa) V_\kappa\psi'
-(\lambda_\kappa-{\bar\lambda}_\kappa)^2 
V_\kappa G_0^{\lambda_\kappa}U^*_{\lambda_\kappa}
\Gamma^{{{\bar \lambda}_\kappa}}T\psi'\non \\
&=& (H_0-{{\bar \lambda}_\kappa})\psi' 
-(\lambda_\kappa-{\bar\lambda}_\kappa)U^*_{{{\bar\lambda}_\kappa}}
\Gamma^{{{\bar \lambda}_\kappa}}T\psi'
+(\lambda_\kappa-{\bar\lambda}_\kappa) V_\kappa\psi'
\non \\
& &\ \ \ \ \ \ -(\lambda_\kappa-{\bar\lambda}_\kappa)^2 
U^*_{{{\bar \lambda}_\kappa}}\Gamma^{\lambda_\kappa}
U_{\lambda_\kappa}U^*_{\lambda_\kappa}
\Gamma^{{{\bar \lambda}_\kappa}}T\psi'\non \\
&=& (H_0-{{\bar \lambda}_\kappa})\psi' 
-(\lambda_\kappa-{\bar\lambda}_\kappa)U^*_{{{\bar\lambda}_\kappa}}
\Gamma^{{{\bar \lambda}_\kappa}}T\psi'
+(\lambda_\kappa-{\bar\lambda}_\kappa) V_\kappa\psi'
\non \\
& &\ \ \ \ \ \ -(\lambda_\kappa-{\bar\lambda}_\kappa) 
U^*_{{{\bar \lambda}_\kappa}}\Gamma^{\lambda_\kappa}
(M^{{{\bar \lambda}_\kappa}}-M^{\lambda_\kappa})
\Gamma^{{{\bar \lambda}_\kappa}}T\psi' \non \\
&= &(H_0-{{\bar\lambda}_\kappa})\psi'. \non
\eea
Therefore $\Ran (H-{{\bar \lambda}_\kappa})$ is dense in ${\cal H}$ and 
$H$ is essentially self-adjoint on $D(H)$.
\par
For $\lambda \in \CC\setminus \NN_0$ such that $0\notin \sigma(M^\lambda)$,  
define
\be
G^\lambda\equiv G^\lambda_0 + U_{\bar \lambda}^* \Gamma^\lambda U_\lambda.
\ee  
One can check using the resolvent identity and identity (\ref{b.9}) that (\cite{G}, see also \cite{AGHKH})
\be
G^\lambda(H-\lambda)\phi =\phi,
\ee
so that 
\be
G^\lambda=(H-\lambda)^{-1}.
\ee  
We now state the main theorem of this paper. (a) is proved in Lemma 3.2,
(c) in Lemma 3.1 and (b) and (d) in Theorem 5.8.
\par\noindent 
{\bf Theorem 2.2}\ {\sl 
\newline
(a)\ The spectrum of $H$ contains bands around the Landau levels $\NN_0$ 
and an interval extending from $-\infty$ to a finite negative point.
\newline 
For each $N\in\NN$ there exists $\kappa_0 > 0$ 
such that for $\kappa >\kappa_0$, with probability one, 
\newline
(b)\ $\sigma_{{\rm cont}}(H)\cap (-\infty, N)=\emptyset $,
\newline
(c)\ if $m \in \NN_0\cap (-\infty,N)$, then $m$
is an eigenvalue of $H$ with infinite multiplicity
\newline 
(d)\ if $\lambda \in \sigma(H)\cap (-\infty,N)\setminus \NN_0$, 
is an eigenvalue of $H$ and the corresponding eigenfunction is 
$\phi_\lambda $, then for any compact subset $B$ of $\CC$, 
$\int_B|\phi_\lambda(z-z')|^2dz'$ decays exponentially in 
$z$ with exponential length less than or equal to $ 2/\kappa$.}
%
\section{The Spectrum}
In this section we study the spectrum of the Hamiltonian.
We first show that the Landau levels are still infinitely degenerate
eigenvalues. We then prove that the spectrum contains bands around the 
Landau levels and an infinite interval in the negative half-line. 
\par
Let $\{U_z: \ z \in \CC\}$ be the family of unitary operators on ${\cal H}$ 
corresponding to the magnetic translations:
\be
\(U_z f\)\(z'\)  = e^{2i\kappa z  \wedge z'} f \(z+z'\).
\label{c.1}
\ee
These satisfy $U_{z_1}U_{z_2} = e^{2i\kappa z_2  \wedge z_1}U_{z_1+z_2}$.   
For $n\in \ZZ[i]$ 
\be
U_n G^\lambda\(\omega\) U_n^{-1}  = G^\lambda\(\tau_n\omega\).
\label{c.2}
\ee
The ergodicity of $\{\tau_m\ :\ m\in\ZZ[i]\}$ and equation (\ref{c.2}) together 
imply that the spectrum of $H(\omega)$ and its components are
non random (see for example \cite{CL}, Theorem V.2.4). We shall first prove that  
almost surely the lower Landau levels are infinitely degenerate eigenvalues
for large $\kappa$. This lemma is a generalization of similar results in 
\cite{DMP1} and \cite{ARB}. The main idea of the proof is to construct 
states in ${\cal H}_m$ which vanish at all the impurity sites, so that they 
are also eigenfunctions of $H$. These states involve the entire function in 
(\ref{c.3}) which vanishes at all the points of $\ZZ[i]$ and consequently 
grows like $e^{A\vert z \vert^2}$ for large $|z|$. The condition that the 
states are square integrable then requires that the magnetic field
be sufficiently large in order to compensate this growth by the factor
$e^{-\kappa \vert z \vert^2}$.
\vskip .25 cm \noindent
{\bf Lemma 3.1} \ {\sl  For each $N\in\NN$, there exists $\kappa_0(N) > 
0$, such that for $\kappa >\kappa_0$, with probability one, 
each Landau level $m$, with $m\leq N$, is infinitely degenerate.}
\par \noindent
{\bf Proof:} \ 
The elements of the space ${\cal H}_0$ are of the form
\be
\phi(z)=\psi(z)e^{-\kappa \vert z \vert^2},
\ee
where $\psi$ is an entire function and, of course, $\phi\in L^2(\CC)$. 
Let 
\be
\psi_0 (z) = z\!\!\!\!\!\!\prod_{n\in\ZZ[i]\setminus\{0\}}\!\!\!\!\! 
(1 - {z \over n})e^{{z\over n} +{{z^2}\over {2n^2}}}.
\label{c.3}
\ee
Then $\psi_0$ is an entire function with zeros at all the points of $\ZZ[i]$.
It follows from the theory of entire functions (see \cite{B} 2.10.1)
that there exists $A>0$ such that
$\vert \psi_0(z)\vert \leq e^{A\vert z \vert^2}$.
For $k \in \NN_0$, let 
\be
\phi_{0,k} (z) 
= z^k\psi_0 (z) e^{-\kappa \vert z \vert^2},
\ee
then, if $\kappa >A$, $\phi_{0,k} \in {\cal H}_0$
and since $V_\kappa\phi_{0,k}=0$, $H\phi_{0,k}=0$.
Also if for $M\in \NN_0$, $\sum^M_{k=0} b_k \phi_{0,k} = 0$, then
$\sum^M_{k=0} b_k z^k = 0$ for $ z \notin \ZZ[i]$.
Therefore $\sum^M_{k=0} b_k z^k \equiv 0$ and thus the $ b_k$'s
are zero implying that the $\phi_{0,k}$'s are linearly independent. 
So the $\phi_{0,k}$'s form an infinite linearly independent set 
of eigenfunctions of $H$ with eigenvalue $0$.  
For the higher levels we modify this argument with the use of the creation and anihilation operators 
for the Hamiltonian $H_0$, $a^*$ and $a$, defined by
$\displaystyle{
a^*=(1/\sqrt {2\kappa})\(-{{\partial\phantom z}\over{\partial z}}+\kappa 
\bar z\)}$ and 
$\displaystyle
{a=(1/\sqrt{2\kappa})\({{\partial\phantom z}\over{\partial \bar z}}+\kappa 
z\)}$. \hfill\break
These operators satisfy the commutation relation $[a,a^*]=1$. 
Also if $\phi \in {\cal H}_m$ then $a^*\phi \in {\cal H}_{m+1}$
and $a\phi \in {\cal H}_{m-1}$ except when $m=0$, in which case $a\phi=0$.
For $m\leq N$ and $k \in \NN_0$, let 
\be
{\tilde\phi}_{m,k} (z) 
= z^k\(\psi_0 (z)\)^{m+1} e^{-\kappa \vert z \vert^2},
\ee
then, if $\kappa >A(N+1)$, 
${\tilde \phi}_{m,k} \in {\cal H}_0$.
Now let $\phi_{m,k}=(a^*)^m{\tilde \phi}_{m,k}$. 
Then $\phi_{m,k} \in {\cal H}_m$ and 
$\phi_{m,k} (n)= 0$ for all $n\in\ZZ[i] $ since ${\tilde \phi}_{m,k}$
has a zero of order greater then $m$ at each point of $\ZZ[i]$.  
Therefore since $V_\kappa\phi_{m,k}=0$, $H\phi_{m,k}=m\phi_{m,k}$.
Moreover since $[a,a^*]=1$ and $a{\tilde \phi}_{m,k}=0$,
$a^m\phi_{m,k}=m!{\tilde \phi}_{m,k}$.
So, if for $M\in \NN_0$, $\sum^M_{k=0} b_k \phi_{m,k} = 0$, then
\be
\sum^M_{k=0} b_k {\tilde \phi}_{m, k} = 
(m!)^{-1}a^m\left(\sum^M_{k=0} b_k \phi_{m,k}\right) = 0.
\ee
This means that
$\sum^M_{k=0} b_k z^k = 0$ for $ z \notin \ZZ[i]$ and as for $m=0$
it follows that the $\phi_{m,k}$'s form an infinite linearly 
independent set of eigenfunctions of $H$ with eigenvalue
$m$.              
\qed
\par
In the case of one impurity of strength $\omega$ at the origin, the Green's function is given by 
\be
G^\lambda= G^\lambda_0 +{1\over {c^\lambda}} G^\lambda_0(\cdot,0) G^\lambda_0(0,\cdot),
\ee
where
\be
c^\lambda={{2\kappa}\over \pi}\(\psi(-\lambda)-{{2\pi}\over \omega}\).
\ee
It is clear that in this case the spectrum consists of Landau levels
and the values of $\lambda$ for which $c^\lambda=0$.
For small $\omega$ the latter correspond to 
points close to the Landau levels and in the case of $\omega>0$, 
there is another point which is negative and of the order of 
$\exp(2\pi/|\omega|)$.
\begin{figure}[hbt] 
\begin{center}\includegraphics[width=10cm]{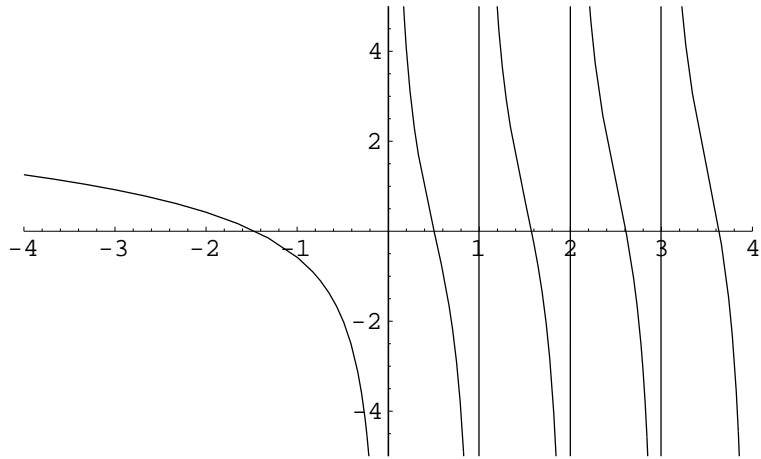}
\end{center}
\caption{$\lambda \mapsto \psi(-\lambda)$}
\label{ }
\end{figure}
In the next lemma we shall 
show that in our case 
these points are also in the spectrum in 
the sense that the spectrum of our Hamiltonian contains
bands around the 
Landau levels and an interval extending 
from $-\infty$ to a finite negative point.  
\par
Let $Y=\{2\pi/ x \ :\ x\in X\setminus \{ 0\}\}$.
\vskip .25 cm \noindent
{\bf Lemma 3.2} \ { \sl With probability one}
\be
-\psi^{-1}\(Y\) \subset \sigma (H (\omega)).
\ee
\par \noindent
{\bf Proof:} \ It is sufficient to prove that for each $\lambda \in 
-\psi^{-1}\(Y\)$ and for 
all $\epsilon> 0$, there exists $\Omega'$ with $\PP (\Omega') > 0$ and 
$\psi \in {\cal H}$ with $\Vert \psi \Vert = 1$ such that for all $\omega \in 
\Omega'$, $\Vert \(G^{\lambda_\kappa}(\omega) - 
\(\lambda-{\lambda_\kappa}\)^{-1}\) \psi \Vert < \epsilon$. 
Let $\la v \vert n \ra =\delta_{n 0}$ and 
let $\psi=C U^*_{\bar \lambda}v$, 
where $C^{-2}= (2\kappa/\pi)\sum_{m=0}^\infty (m-\lambda)^{-2}$.
Note that $\psi(z)=C G^\lambda_0(z,0)$ and
$\Vert \psi \Vert=1$ by (\ref{b.10}).
Then
\bea
\bigl (G^{\lambda_\kappa} - 
\!\!\!\!\!\!\!&&\!\!\!\!\!\!\!
\(\lambda-{\lambda_\kappa}\)^{-1}\bigr ) \psi \non \\
&=& 
\(G^{\lambda_\kappa}_0 
+ U^*_{\bar{\lambda}_\kappa}\Gamma^{\lambda_\kappa}U_{\lambda_\kappa}
+ \({\lambda_\kappa}-\lambda\)^{-1}\) \psi \non \\
&=& 
\({\lambda_\kappa}-\lambda\)^{-1}
C\(\({\lambda_\kappa}-\lambda\)
G^{\lambda_\kappa}_0 U^*_{\bar \lambda} 
- U^*_{\bar{\lambda}_\kappa}\Gamma^{\lambda_\kappa}
(M^{\lambda_\kappa}-M^\lambda)
+U^*_{\bar \lambda}\) v \non \\
&=& 
\({\lambda_\kappa}-\lambda\)^{-1}
C\(U^*_{\bar{\lambda}_\kappa}-U^*_{\bar \lambda} 
- U^*_{\bar{\lambda}_\kappa}
+ U^*_{\bar{\lambda}_\kappa}\Gamma^{\lambda_\kappa}
M^\lambda+U^*_{\bar \lambda}\) v \non \\
&=&\({\lambda_\kappa}-\lambda\)^{-1}
CU^*_{\bar{\lambda}_\kappa}\Gamma^{\lambda_\kappa}M^\lambda v.\non
\eea
By using (\ref{b.9}) we get 
\bea
\Vert \(G^{\lambda_\kappa}(\omega) - 
\(\lambda-{\lambda_\kappa}\)^{-1}\) \psi \Vert^2
&=& C^2\vert \lambda-{\lambda_\kappa}\vert^{-2} \({\cal I}\lambda_\kappa\)^{-1} 
{\cal I}\la M^\lambda v, \Gamma^{\lambda_\kappa}M^\lambda v\ra\non \\
&\leq&
2C^2 \vert \lambda-{\lambda_\kappa}\vert^{-2} 
\vert {\cal I}\lambda_\kappa\vert^{-1} 
\Vert M^\lambda v\Vert\ 
\Vert {\tilde\Gamma}^{{\bar\lambda}_\kappa}M^\lambda v\Vert \non
\eea
by (\ref{b.6}).
Choose $R$ such 
that $\sum_{\vert n \vert > R} \vert G^{\lambda}_0(n,0)\vert^2 
< \delta$ 
and let 
\be
\Omega' = \{\omega: \vert c^\lambda_0\vert < \delta,
\ \ \min_{\vert n\vert \leq R, n \neq 0}
\vert c^{\lambda_\kappa}_n \vert > 1/\delta \}.
\ee 
Since $\psi(-\lambda)\in Y$ and $0$ is in the support of $\mu$, 
$\PP (\Omega') > 0$. 
We have
\be
\la n\vert M^\lambda v\ra = \cases { c^\lambda_0, & if $n=0$ \cr
-G^\lambda_0(n,0), & if $n\neq 0$.\cr}
\ee 
Therefore
\be
\Vert M^\lambda v\Vert^2 \leq
\delta^2 + \sum_{n \neq 0} \vert G^{\lambda}_0(n,0)\vert^2 
\ee       
and
\bea
\Vert {\tilde\Gamma}^{{\bar\lambda}_\kappa} M^\lambda v\Vert^2
&=& \vert c_0^\lambda\vert^2\vert c_0^{\lambda_\kappa}\vert^{-2}
+ \sum_{n \neq 0} \vert c_n^{\lambda_\kappa}\vert^{-2}
\vert G^{\lambda}_0(n,0)\vert^2 \non \\
&\leq& \delta^2(\pi/2\kappa)^2 \vert {\cal I}\psi(-\lambda_\kappa)\vert^{-2}
+ \delta^2 \sum_{\vert n \vert \leq R}\vert G^{\lambda}_0(n,0)\vert^2 \non \\
&&\ \ \ \ \ \ \ \ \ \ \ \ \ \ \ \ \ \ \ \ \ \ \ \ \ \ \ \ \ \ \ \ \ \ \ \ 
+ (\pi/2\kappa)^2 \vert {\cal I}\psi(-\lambda_\kappa)\vert^{-2}
\sum_{\vert n \vert >R}\vert G^{\lambda}_0(n,0)\vert^2 \non \\
&\leq &\delta^2(\pi/2\kappa)^2 \vert {\cal I}\psi(-\lambda_\kappa)\vert^{-2}
+ \delta^2 \sum_{n \neq 0}\vert G^{\lambda}_0(n,0)\vert^2 \non \\
&&\ \ \ \ \ \ \ \ \ \ \ \ \ \ \ \ \ \ \ \ \ \ \ \ \ \ \ \ \ \ \ \ \ \ \ \ 
+ \delta(\pi/2\kappa)^2 \vert {\cal I}\psi(-\lambda_\kappa)\vert^{-2}. \non
\eea       
Thus 
$\Vert \(G^{\lambda_\kappa}(\omega) - 
\(\lambda-{\lambda_\kappa}\)^{-1}\) \psi \Vert<\epsilon $
if $\delta$ is small enough.
\qed 
\par
In the next section we relate the generalized eigenvectors of $H$ with those of $M^\lambda$.
%
\section{  Generalized eigenfunctions of $H$}
In this section we show that a generalized eigenfunction of $H$ with eigenvalue
$\lambda$, say, which is not a Landau level, is related in a simple way to an eigenvector $v$ of $M^{\lambda}$ with eigenvalue zero. Furthermore if $v$ decays then so does the corresponding eigenfunction. Since this reduces the problem to a lattice problem, it makes it possible for us to use the Aizenman-Molchanov method.
\par\noindent
{\bf Proposition 4.1}\  \ {\sl If $\phi$ is a generalized eigenfunction of $H$ 
with eigenvalue $\lambda\notin \NN_0$, 
then $v=\Gamma^{\lambda_\kappa} U_{\lambda_\kappa}\phi$
is a generalized eigenvector of $M^{\lambda}$ with eigenvalue zero and
$\phi=(\lambda-{\lambda_\kappa})U^*_{\lambda}v$.
Moreover if $v$ decays exponentially, then for any compact subset 
$B$ of $\CC$, $\int_B\vert \phi(z-z')\vert^2 dz'$ decays exponentially in $z$.}
\par\noindent
{\bf Proof:}\ 
Suppose $\phi$ is a generalized eigenvector of $H$ with eigenvalue $\lambda$. Then
\be
G^{\lambda_\kappa} \phi = (\lambda - {\lambda_\kappa})^{-1} \phi  \label{d.1} 
\ee
or
\be
G_0^{\lambda_\kappa} \phi + U_{\bar {\lambda_\kappa}}^* \Gamma^{\lambda_\kappa} U_{\lambda_\kappa}\phi
= (\lambda - {\lambda_\kappa})^{-1} \phi.  \label{d.2} 
\ee
Thus
\be
U_{\lambda} G_0^{\lambda_\kappa} \phi + 
U_{\lambda} U_{\bar {\lambda_\kappa}}^* \Gamma^{\lambda_\kappa} U_{\lambda_\kappa}\phi
= (\lambda - {\lambda_\kappa})^{-1} U_{\lambda}\phi.  \label{d.3} 
\ee
Using $U_{\lambda} G_0^{\lambda_\kappa} = (\lambda - {\lambda_\kappa})^{-1} 
(U_{\lambda} - U_{\lambda_\kappa})$, we get
\be
U_{\lambda} U_{\bar{\lambda_\kappa}}^* \Gamma^{\lambda_\kappa} U_{\lambda_\kappa}\phi
= (\lambda - {\lambda_\kappa})^{-1} U_{\lambda_\kappa}\phi  \label{d.4} 
\ee
which by (\ref{b.9}) can be written in the form
\be
M^{\lambda} \Gamma^{\lambda_\kappa} U_{\lambda_\kappa}\phi = 0.
\ee
Therefore if $v=\Gamma^{\lambda_\kappa} U_{\lambda_\kappa}\phi$,
\be
M^{\lambda} v = 0.
\ee
From (\ref{d.2}) we get
\be
(\lambda - {\lambda_\kappa})G_0^{\lambda_\kappa} \phi 
+ (\lambda - {\lambda_\kappa})U_{\bar {\lambda_\kappa}}^* v
= \phi.  \label{d.5} 
\ee
Thus
\be
(\lambda - {\lambda_\kappa})G_0^{\lambda}G_0^{\lambda_\kappa}  \phi 
+ (\lambda - {\lambda_\kappa})G_0^{\lambda}U_{\bar {\lambda_\kappa}}^* v
= G_0^{\lambda}\phi.  
\label{d.6} 
\ee
By using the resolvent identity we can write this as
\be
U_{\lambda}^* v
=G_0^{\lambda_\kappa} \phi + U_{\bar {\lambda_\kappa}}^* v
\label{d.7} 
\ee
and therefore $\phi=(\lambda-{\lambda_\kappa})U^*_{\lambda}v $ by 
(\ref{d.5}). 
From Propositions 6.1 and 6.2 in Appendix A we get for
$\lambda\notin \NN_0$
\be
\vert G^{\lambda}_0 (z,z') \vert < C 
e^{- {\kappa \over 2} \vert z - z' \vert^2}
(1 + {\bf 1}_{B(0, 1/\sqrt{2 \kappa})} (\vert z - z' \vert) 
\vert \ln \vert z - z' \vert \vert)
\ee
where $C$ depends on $\lambda$ and $\kappa$.
From the equation
\be
\phi (z) = (\lambda - {\lambda_\kappa}) \sum_{n \in \ZZ  [i]} G_0^{\lambda} (z,n) 
\la n \vert v\ra,
\ee
we get, assuming $\la n \vert v\ra \leq C' e^{- \alpha \vert n 
\vert}$, that
\bea
\vert \phi (z) \vert 
& \leq &\vert \lambda - {\lambda_\kappa} \vert C' 
\sum_{n \in \ZZ [i]} \vert G_0^{\lambda} (z, n) \vert 
e^{- \alpha \vert n \vert} \non \\
& \leq & C'' \sum_{n \in \ZZ [i]} e^{- {\kappa \over 2} \vert z  - n \vert^2} 
e^{- \alpha \vert n \vert} \non \\
& + & C'' \sum_{n \in \ZZ [i]} e^{- {\kappa \over 2} \vert z - n \vert^2} 
e^{- \alpha \vert n \vert} {\bf 1}_{B(0, 1/\sqrt{2 \kappa})} 
(\vert z  - n \vert) \vert \ln \vert z - n \vert \vert \non \\
& = & S_1 + S_2. \non 
\eea
Now
\bea
S_1 
& \leq & C'' \sum_{\vert n - z \vert \geq 1} 
e^{- {\kappa \over 2} \vert z - n \vert} e^{- \alpha \vert n \vert} 
+ C''  e^\alpha e^{- \alpha \vert z \vert} \non \\
& \leq & C'' e^{- \beta \vert z \vert} \sum_{n \in \ZZ [i]} 
e^{- \beta \vert n \vert} + C'' e^\alpha e^{- \alpha \vert z \vert} \non 
\eea
where $\beta = {1 \over 4}\min(\kappa, 2 \alpha)$.
Thus, $S_1 \leq C''' e^{- \beta \vert z \vert}$.
Similarly
\be
S_2 \leq C''' e^{- \beta \vert z \vert} \sum_{n \in \ZZ [i]} 
e^{- \beta \vert n \vert}{\bf 1}_{B(0,{1 \over {\sqrt{2 \kappa}}})} 
(\vert z - n \vert)\ln \vert z - n \vert \vert.
\ee
Therefore
\be
\vert \phi (z) \vert^2 \leq C e^{- 2 \beta \vert z \vert} (1 + 3 \sum_{n \in 
\ZZ [i]} e^{- \beta \vert n \vert}
{\bf 1}_{B(0, 1/\sqrt{2 \kappa})} (\vert z - n  \vert) 
\vert \ln \vert z - n \vert \vert^2).
\ee
Let $B \subset \CC$ be compact and let $R = 
\sup \{\vert z \vert: z \in B\}$. Then for $z' \in B$
\be
\vert \phi (z - z') \vert^2 \leq C e^{2 \beta R} e^{- 2 \beta \vert z 
\vert} 
(1 + 3 \sum_{n \in \ZZ [i]} e^{- \beta \vert n \vert}
{\bf 1}_{B (0, {1 \over {\sqrt{2 \kappa}}})} 
(\vert z - z' - n \vert) \vert \ln \vert z - z' - n \vert \vert^2).
\ee
Therefore
\be
\int_B\vert \phi (z - z') \vert^2 dz'
\leq C e^{2 \beta R} e^{- 2 \beta \vert z \vert} 
(\vert B \vert + 3 \sum_{n \in \ZZ [i]} e^{- \beta \vert n \vert}
\int_{\vert z' \vert < {1\over {\sqrt{2 \kappa}}}} 
\vert \ln \vert z' \vert \vert^2).
\ee
\qed
\par
We do not dwell on the existence of the generalized eigenfunctions. It suffices to say that the arguments of Theorem II.4.5 in \cite{CL} can be used with $e^{-tH}$ replaced by $G^{\lambda_\kappa}$ since from the bound in Lemma 2.1 and the bounds in Appendix A for $|G^\lambda_0(z,z')|$ it follows that
\be
\sup_z \int_{\CC}|G^{\lambda_\kappa}_0(z,z')|^2 dz'<\infty.
\ee
The same bounds guarantee also that $v$ is a generalized eigenvector of $M^\lambda$. In the next section we apply the Aizenman-Molchanov method to the lattice operator $M^\lambda$.
%
\section{An Application of the Aizenman-Molchanov Method}
In this section we apply the Aizenman-Molchanov method to $M^\lambda$,
where $\lambda$ is not a Landau level. The main ingredient in this method is 
the Decoupling Principle for $\tau$-regular measures. We start by stating this principle, not in its full generality but in the form in which it will be used here.
\par\noindent
{\bf Definition} {\sl A measure $\mu$ on $\RR$ is said to be $\tau$-regular,
with $\tau\in (0,1]$, if there exists $\nu >0$ and $C<\infty$ such that 
\be
\mu([x - \delta, \ x + \delta]) \leq C \delta^\tau \mu([x - \nu, x+\nu])
\ee
for all $x \in \RR$ and $0 < \delta < 1$.}
\par\noindent
{\bf Lemma 5.1 (A Decoupling Principle)}
{\sl
Let $\mu$ be a $\tau$-regular measure and let $\int |u |^\epsilon 
\mu (du) < \infty$ for some $\epsilon >0$. Then for all $0<s < \min(\tau, \epsilon)$
there exists $\xi_s $, a positive, increasing function on $\RR_+$ 
with $\xi_s(0)>0 $ satisfying 
\be
\lim_{x\to\infty} {{\xi_s (x)} \over x} =  1,
\ee
such that for all $\eta$, $a$ and $b\in\CC$,
\be
\int |u - \eta |^s |au+b|^{-s} \mu (du) 
\geq 
(\xi_s ( |\eta |))^s \int |au+b |^{-s} \mu (du).
\label{e.1}
\ee
}
\noindent
Let $\mu (A)=\mu_0(\{\omega\ :\ 1/\omega \in A\})$.
In Appendix B we shall show that $\mu$ is 1-regular and 
$\int |u  |^\epsilon \mu (du) < \infty$ for all $\epsilon <1$.
Thus the inequality ({\ref{e.1}) is valid for all 
$s\in (0, 1)$. As in \cite{AM} we use this lemma to obtain an exponential
bound on $\la n | \Gamma^\lambda (z) |0\ra $ where
$\Gamma^\lambda (z)= (M^\lambda - z)^{-1}$. This bound then allows us to apply the results of \cite{SW} to deduce that the spectrum of $M^\lambda$ in a neighbourhood of the origin consists of eigenvalues and that the corresponding eigenvectors decay exponentially. We then combine this result with Proposition 4.1 to translate it into a statement about the properties of the spectrum of $H$.
\par 
It is convenient here to introduce a notation for the intervals between the Landau levels. We let $I_0=(-\infty, 0)$ and $I_N =(N-1, N)$ for $N\in \NN$.
\par\noindent
{\bf Lemma 5.2}\ 
{\sl For all $N\in \NN_0$, for all $s\in (\half, 1)$
and for all $\gamma < s$
there exists $\kappa_0 (N,s) < \infty$ such that for all $\kappa > \kappa_0 (N,s)$,
for all $\lambda \in (-\infty, N)\setminus \NN_0$ and for all $z\in \CC$ with 
${\cal I}z\neq 0$ and $|{\cal R}z |\leq 1$, 
\be
\EE \{\sum_{n \in \ZZ [i]} |\la   n |\Gamma^\lambda (z) 
|0 \ra |^s e^{\gamma\kappa|n |} \} \leq 
1/\{2\kappa(\xi_s(0))^s\}.
\label{e.2}
\ee   
}
{\bf Proof:}\  The starting point is the following equation: For $z\notin \RR$,
\be
\sum_{n' \in \ZZ [i]} \la n |M^\lambda - z |n' \ra \la n' 
|\Gamma^\lambda (z) |n'' \ra = \delta_{nn''}.\ee
This becomes using (\ref{b.3})
\be
(c^\lambda_n - z) \la n |\Gamma^\lambda (z) |n'' 
\ra - \sum_{n' \not= n} G_0^\lambda (n, n') \la n' |\Gamma^\lambda (z)
|n'' \ra = \delta_{nn''}.
\ee
Now we take $n \not= n''$ and $0 < s < 1$ to get
\bea
|c_n^\lambda  - z |^s |\la n |
\Gamma^\lambda  (z) |n'' \ra |^s 
& = &|\sum_{n' \not= n} G_0^\lambda (n,n') \la n' |\Gamma^\lambda 
(z) |n'' \ra |^s \non \\
& \leq &\sum_{n'\not= n} |G_0^\lambda (n, n') |^s 
|\la n' |\Gamma^\lambda (z) |n'' \ra |^s. \non 
\eea
Thus
\be\EE \{ |c^\lambda_n - z |^s |\la n |\Gamma^\lambda 
(z) |n'' \ra |^s \}
\leq \sum_{n' \not= n} |G_0^\lambda (n,n') |^s \EE \{ |\la n' 
|\Gamma^\lambda (z) |n'' \ra |^s\}.
\label{e.2b}
\ee
Now
\be
\EE \{ |c_n^\lambda - z |^s |\la n |\Gamma^\lambda 
(z) |n'' \ra |^s \}
=
{\tilde {\EE}}_n \EE_n \{|c_n^\lambda - z |^s |\la n 
|\Gamma^\lambda (z) |n'' \ra |^s\}
\ee
where $\EE_n$ is the expectation with respect to $\omega_n$ and ${\tilde 
{\EE}}_n$ is with respect to all other $\omega_{n'}$'s.
Let
\be
\la n' |M_n^\lambda  |n'' \ra = \la n' |M^\lambda |n'' 
\ra - (4\kappa/\omega_n) \delta_{nn'} \delta_{nn''}.
\ee
Then  $M^\lambda_n$ is independent of $\omega_n$ and using the resolvent identity 
\be
\la n |\Gamma^\lambda (z) |0 \ra 
= {A \over {1 + (4\kappa/\omega_n) B}}
\ee
where
$A= \la n|(M_n^\lambda - z)^{-1} |0 \ra$
and 
$B= \la n |(M_n^\lambda - z)^{-1} |n \ra$.
Then 
\bea
\EE_n \{|c_n^\lambda - z |^s 
\!\!\!\!\!\! &| &\!\!\!\!\!\!\la n |
\Gamma^\lambda (z) |0 \ra |^s \} \non \\
& = & 
\EE_n \left\{{{|c_n^\lambda - z |^s} 
\over {|1 + (4\kappa/\omega_n)B |^s}}  \right\} |A |^s \non \\
& \geq &(4\kappa)^s \int {{|u -\eta|^s} \over 
{|1+4\kappa u B|^s}} \mu (du) |A |^s \non 
\eea
where $u=1/\omega_n$ and
$2\pi \eta= \psi(- \lambda)-{{\pi} \over {2 \kappa}} E 
$,
$E$ being the real part of $z$.
Thus by Lemma 5.1
\bea
\EE_n \{|c_n^\lambda - z |^s |\la n |
\Gamma^\lambda (z) |0 \ra |^s \} 
& \geq &(4\kappa)^s
(\xi_s (|\eta |))^s \EE_n {{|A |^s} \over 
{|1 + c_n^\lambda B|^s}} \non \\
& = &(4\kappa)^s
(\xi_s (|\eta|))^s \EE_n \{|\la n |
\Gamma^\lambda (z) |0 \ra |^s\} \non 
\eea
Using (\ref{e.2b}) this gives
\be
(4\kappa)^s(\xi_s (|\eta |))^s \EE \{ |\la n |
\Gamma^\lambda (z) |0 \ra |^s \} 
\leq \sum_{n' \not= n} |G_0^\lambda (n,n') |^s \EE \{|\la n' 
|\Gamma^\lambda (z) |0  \ra |^s \}.
\ee
or
\be
\EE \{ |\la n |\Gamma^\lambda (z) |0 \ra |^s \} 
\leq(1/4\kappa)^s(\xi_s (|\eta|))^{-s}
\sum_{n' \not= n} |G_0^\lambda (n, n') |^s \EE \{ |\la 
n' |\Gamma^\lambda (z) |0 \ra |^s \}. 
\ee
Let $\gamma>0$ and define
\be
\Xi (s) 
= \EE\{\sum_{n \in \ZZ [i]} e^{{\gamma\kappa} |n |} |\la n 
|\Gamma^\lambda (z) |0 \ra |^s \}.
\ee
Then 
\bea
\Xi (s) 
&=& \EE \{|\la 0 |\Gamma^\lambda (z) |0 \ra 
|^s \}
+ \sum_{n \not= 0} e^{{\gamma\kappa} |n |} \EE\{|\la n |\Gamma^\lambda 
(z) |0 \ra |^s \}\non \\
&\leq &\EE \{|\la 0 |\Gamma^\lambda (z) |0 \ra |^s 
\}\non \\
& &
\ \ \ \ + (1/4\kappa)^s(\xi_s (|\eta|))^{-s} 
\sum_{n \not= 0} 
\sum_{n' \not= n}
e^{{\gamma\kappa} |n |} |G_0^\lambda (n,n') |^s \EE\{|\la n' |
\Gamma^\lambda (z) |0 \ra |^s\}.\non 
\eea
Thus   
\bea
\Xi (s) 
&\leq &\EE\{|\la 0 |\Gamma^\lambda (z) |0 
\ra |^s\}\non \\
& &
+ (1/4\kappa)^s
(\xi_s (|\eta |))^{-s} \sum_{n'} \sum_{n\neq n'}
\!\!\!\! e^{{\gamma\kappa} |n - n' |} 
|G_0^\lambda (n,n') |^s e^{\gamma\kappa 
|n' |}
\EE \{|\la n'|\Gamma^\lambda (z) |0 
\ra |^s\}\non 
\eea
so that
\be
\Xi (s) \leq \EE \{ |\la 0 |\Gamma^\lambda (z) |0 
\ra |^s\}
+ (1/4\kappa)^s
\sup_{n'} {{\sum_{n \neq n'}e^{{\gamma\kappa} |n - n' |} |G_0^\lambda 
(n,n') |^s} \over {(\xi_s |\eta |))^s}} 
\Xi (s).\ee      
Let
\be
F(s,\lambda) = (1/4\kappa)^s
\sup_{n'} {{\sum_{n \not= n'} e^{{\gamma\kappa} |n - n' |} 
|G_0^\lambda (n,n') |^s} \over 
{\left(\xi_s (|\eta |)\right)^s}}.
\ee
If $F (s,\lambda) < 1/2$ then
\be
\Xi (s) \leq {{\EE\{|\la 0 |\Gamma^\lambda (z) |0 
\ra |^s\}} \over {1 - F (s, \lambda)}}
\leq 2\EE\{|\la 0 |\Gamma^\lambda (z) |0 
\ra |^s\}.
\ee
But
\be
\la 0 |\Gamma^\lambda (z) |0 \ra 
= {{\la 0 |(M_0^\lambda - z)^{-1} |0 \ra} 
\over {1 + c_0^\lambda \la 0 
|(M_0^\lambda - z)^{-1} |0 \ra}} 
\ee
so that
\be
|\la 0 |\Gamma^\lambda (z) |0 \ra |
= {1 \over {4\kappa|b+1/\omega_0 |}} 
\ee
where $b$ is independent of $\omega_0$.
Using this and Lemma 5.1 with $a=1$ and $\eta=-b$, we get
\be
\EE_0\{|\la 0 |\Gamma^\lambda (z) |0 \ra |^s\}\leq  
1/(4\kappa \xi_s(0))^s
\ee
and therefore
\be
\EE\{| \la 0 |\Gamma^\lambda (z) |0 \ra |^s\} \leq 
1/(4\kappa \xi_s(0))^s.
\ee
This proves (\ref{e.2}). To prove that $F (s,\lambda) < 1/2$,
First assume that $ \lambda \in  I_N$ with $N\in \NN$.   By Proposition 6.2, we have
\be
|G_0^\lambda (n,n') |\leq  (C \kappa)^{N +1}N^N
|\Gamma (- \lambda) ||n - n' |^{2N} e^{- \kappa |n - n'|^2}
\ee
for $n \not= n'$, and $ \lambda \in  I_N$, $N\in \NN$.
Therefore  
\be
\sum_{n \not= n'} e^{{\gamma\kappa} |n - n' |} 
|G_0^\lambda (n, n') |^s
\leq (C \kappa)^{(N+1)s} N^{Ns} |\Gamma (- \lambda) |^s
\sum_{n \not= 0} e^{{\gamma\kappa} |n |} 
 |n |^{2Ns} e^{- \kappa s |n |^2}.
\ee
Let $\gamma< \alpha <s $. Using the bounds 
$ e^{{\gamma\kappa}x} e^{- \alpha \kappa  x^2}\leq e^{-(\alpha -
\gamma)\kappa}$ for $x\geq 1$, \break
$ x^{2Ns} e^{- \kappa (s -\alpha)x^2/2} < (2sN/(s-\alpha)\kappa )^{Ns}$ 
for $x\geq 1$ and
\be
\sum_{n \in \ZZ[i]} e^{- t |n |^2}
\leq K(t)
\ee
where $K(t)=\(1+e^{-t/4}+(\pi/t)^{1/2}\)^2 $ for $t>0$ (see Lemma 2.1 in 
\cite{DMP1}), we get
\bea
\sum_{n \not= n'} e^{{\gamma\kappa} |n - n' |}\!\!\!\!\!\!\! 
&&\!\!\!\!\!\!\! |G_0^\lambda (n, n') |^s \non \\
&\leq & \(K(\kappa (s-\alpha)/2) -1\) (C^{N+1}\kappa)^s (2sN^2/(s-\alpha))^{Ns}
|\Gamma (- \lambda) |^s e^{-(\alpha -\gamma)\kappa}.\non 
\eea
Thus
\be
F(s, \lambda) \leq 
\(K\({{\kappa (s-\alpha)}\over 2}\) -1\) (C^{N+1}/4)^s (2sN^2/(s-\alpha))^{Ns}
e^{-(\alpha -\gamma)\kappa}
\left |{{ \Gamma(- \lambda)} 
\over {\xi_s (|\eta |) }}\right|^s.
\ee
Now since for $N\in \NN$, the limits
$\lim_{\lambda \to N}|\Gamma(-\lambda)/\psi(-\lambda) | $
and \hfill \break $\lim_{\lambda \to N-1}|\Gamma(-\lambda)/\psi(-\lambda) | $ are finite, 
we have for $\lambda\in I_N$, 
\be
|\Gamma(-\lambda)|
\leq C_N (1 + |\psi(-\lambda) |)
\leq C_N(1 + (\pi/2\kappa)+2\pi|\eta|).
\ee
Therefore  
\be
\left |{{ \Gamma(- \lambda)} 
\over {\xi_s (|\eta |) }}\right|
\leq C_N\(\[\{1+(\pi/2\kappa)\}/\xi_s (0)\]
+2\pi\sup_{x\in \RR_0}\{x/\xi_s (x)\}\).
\ee
Thus there exists $\kappa (N_1,s) < \infty$ such that 
for all $\kappa > \kappa_0 (N_1,s)$,
$F(s,\lambda)< 1/2$ for all $\lambda\in (0, N_1)\setminus \NN$. 
\par
For $\lambda \in I_0$ we have by Proposition 6.1
\be
|G_0^\lambda (n,n') |\leq  (C \kappa) { 1\over {|\lambda|}}
e^{- \kappa |n - n' |^2}
\ee
and therefore  
\be
\sum_{n \not= n'} e^{{\gamma\kappa} |n - n' |} 
|G_0^\lambda (n, n') |^s
\leq \(K(\kappa (s-\alpha)) -1\) (C\kappa)^s \left ( { 1\over {|\lambda|}} \right )^s  e^{-(\alpha -\gamma)\kappa}
\ee
and thus
\be
F(s, \lambda) \leq 
\(K(\kappa (s-\alpha)) -1\)(C/4)^s e^{-(\alpha -\gamma)\kappa} s \left ( {1\over {|\lambda|\xi_s (|\eta|)}} \right )^s.
\ee
Therefore by the same argument 
$F(s,\lambda)< 1/2$ for all $\lambda\in I_0$ if $\kappa$ is large 
enough. 
\qed
\par
We have from Theorems 8 and 9 in \cite{SW} that if for all $E\in(-1,1)$ and a.e. $\omega$,
\be
\lim_{\epsilon\downarrow 0}
\sum_{n \in\ZZ[i]}|\la n|\Gamma^\lambda(E+i\epsilon)|0\ra|^2 <\infty,
\label{e.3a}
\ee
then $\sigma_{{\rm cont}}(M^\lambda)\cap(-1,1)=\emptyset $ for a.e. $\omega$.
If, furthermore, for a.e. pair $(\omega, E)$, $\omega\in\Omega$ and 
$E\in (-1,1)$,
\be
\lim_{\epsilon\downarrow 0}
|\la n |\Gamma^\lambda(E+i\epsilon)|0\ra|<C_{\omega,E}e^{-m(E)|n|},
\label{e.3b}\ee
then with probability one, the eigenvectors $v^\lambda_E$ of $M^\lambda$ with eigenvalue
$E\in (-1,1)$ obey
\be
|\la v^\lambda_E |n \ra|<D_{\omega,E}e^{-m(E)|n|}.
\label{e.4}
\ee
We shall use the results of \cite{SW}, Lemma 5.2 and 
Proposition 4.1 to prove the  
following lemma.
\par\noindent
{\bf Lemma 5.3}\ {\sl For each $N\in\NN$ there exists $\kappa_0 > 0$ 
such that for $\kappa >\kappa_0$, for each 
$\lambda\in (-\infty,N)\setminus \NN_0 $ with probability one, 
if $\lambda $ is a generalized eigenvalue of $H$ with corresponding generalized eigenfunction  $\phi_\lambda$, then for any compact subset $B$ of $\CC$, 
$\int_B|\phi_\lambda(z-z')|^2dz'$ decays exponentially in 
$z$ with exponential length less than or equal to $ 2/\kappa $.}  
\newline
{\bf Proof:}\ \
From Lemma 5.2 we have for all $\lambda \in(-\infty,N)\setminus \NN_0$, 
$z\in \CC$ with 
${\cal I}z\neq 0$ and $|{\cal R}z |\leq 1$, 
\bea
\EE \Bigl \{\Bigl [\sum_{n \in \ZZ [i]} |\la n |\Gamma^\lambda (z) 
|0 \ra |^2 e^{2\gamma\kappa|n |/s}\Bigr ]^{s/2} \Bigr \} 
&\leq &
\EE \{\sum_{n \in \ZZ [i]} |\la n |\Gamma^\lambda (z) 
|0 \ra |^s e^{\gamma\kappa|n |} \}  \non \\ 
&\leq &
1/\{(2\kappa)(\xi_s(0))^s\}.
\label{e.5}
\eea 
Now for a.e. pair $(\omega, E)$, $\omega\in\Omega$ and 
$E\in (-1,1)$, $ \lim_{\epsilon\downarrow 0}
 \la   n |\Gamma^\lambda (E+i\epsilon) 
|0 \ra $  exists. Therefore by Fatou's Lemma,
\bea
\EE \Bigl\{\Bigl [\sum_{n \in \ZZ [i]} \lim_{\epsilon\downarrow 0}
|\la   n |\Gamma^\lambda (E
\! \! \! \! \! &+&\! \! \! \! \! 
i\epsilon) |0 \ra |^2 e^{2\gamma\kappa|n |/s} \Bigr ]^{s/2}\Bigr \} \non \\
&\leq & \EE \Bigl \{\liminf_{\epsilon\downarrow 0} \Bigl [\sum_{n \in \ZZ [i]} 
|\la   n |\Gamma^\lambda (E+i\epsilon) 
|0 \ra |^2 e^{2\gamma\kappa|n |/s}  \Bigr]^{s/2}\Bigr \} \non \\ 
&\leq &
\liminf_{\epsilon\downarrow 0}\EE \Bigl \{ \Bigl [ \sum_{n \in \ZZ [i]} 
|\la   n |\Gamma^\lambda (E+i\epsilon) 
|0 \ra |^2 e^{2\gamma\kappa|n |/s} \Bigr ]^{s/2} \Bigr \} \non \\ 
&\leq &
1/\{(2\kappa)(\xi_s(0))^s\}.
\label{e.6}
\eea 
Thus (\ref{e.3a}) 
and (\ref{e.3b}) are satisfied. 
Therefore if $\lambda$  
is a generalized eigenvalue of $H$ with corresponding generalized eigenfunction  $\phi_\lambda$, then by Proposition 4.1, $v_\lambda= \Gamma^{\lambda_\kappa}U_{\lambda_\kappa}\phi_\lambda$
is a generalized eigenvector of $M^\lambda$ with eigenvalue $0$
and must satisfy 
\be
|\la v_\lambda |n \ra|<D_{\omega }e^{-2\gamma\kappa|n |/s}.
\label{e.7}
\ee
Then again by Proposition 4.1, for any compact subset $B$ of $\CC$, 
$\int_B|\phi_\lambda(z-z')|^2dz'$ decays exponentially in 
$z$ with exponential length less than or equal to 
\hfill\break
$\max(s/(2\gamma\kappa), 2/\kappa)$. If we choose $\gamma=s/2$, then 
$\max (s/(2\gamma\kappa), 2/\kappa)= 2/\kappa $.\par
\qed
\par
By Fubini's Theorem, we can deduce from Lemma 3.5 the result about the decay of eigenfunctions with probability one and a.e. $\lambda$ with respect to Lebesgue measure and therefore with probability one
$\sigma_{{\rm ac}}(H)\cap (-\infty, N)=\emptyset $. However to be able to make a
statement about $\sigma_{{\rm cont}}(H)$ we have to replace
{\sl a.e. $\lambda$ with respect to Lebesgue measure} with 
{\sl a.e. $\lambda$ with respect to the spectral measure of $H(\omega)$}.
We do this in the Lemma 5.7 by using the ideas of \cite{DLS} and the following four lemmas.
\par\noindent
We state the first lemma without proof.
\vskip .2 cm \noindent
{\bf Lemma 5.4} \ {\sl Let $\{f_n\}$ be a total countable subset of normalized vectors of 
a Hilbert space ${\cal H}$ and $H$ a self-adjoint operator on ${\cal H}$ with 
spectral projections ${\bf E}(\ \cdot \ )$. Let $c_n > 0$, $\sum_n c_n < \infty$ and 
$\nu = \sum_n c_n \mu_n$, where $\mu_n  = (f_n, {\bf E} (\ \cdot \ ) f_n)$. Then 
$\nu(A) = 0$ implies that ${\bf E} (A) = 0$.}
\par\noindent
{\bf Lemma 5.5} \ {\sl For each $N \in \NN_0$, there exists an open set $J_N\subset \CC$, containing $I_N$, such that for $\kappa$ sufficiently large with probability 
one, $M^\lambda$ is invertible for all $\lambda \in J_N \setminus I_N$.}
\par \noindent
{\bf Proof }\ Let $\lambda \in I_N$ and $| \epsilon | < 1$, $\epsilon \neq 0$. 
Let 
\be 
\la n | X | n' \ra = \cases{c_n^{\lambda + i \epsilon} & if $n = 
n'$ \cr 
- G_0^\lambda (n,n') & if $n \neq n'$  \cr}
\ee              
Then $|| X \xi || \geq {{2 \kappa} \over \pi} 
|{\cal I} \psi (- (\lambda + i \epsilon))| ||\xi ||$. 
Therefore $X$ is invertible and
\be
|| X^{-1} || \leq {\pi \over {2 \kappa}} 
{1 \over {|{\cal I}(\psi (- (\lambda + i \epsilon))|}} 
\ee
Let 
\be
\la n | Y | n' \ra = \cases{0 & if $n = n'$ \cr
- i \epsilon (G_0^\lambda G_0^{\lambda + i \epsilon})(n, n') 
& if $n \neq n'$ \cr}
\ee
so that 
\be 
 M^\lambda  = X + Y  =  X (1 + X^{-1} Y).
\ee
From Proposition 6.2 in Appendix A we have for $\lambda $ with ${\cal R} \lambda \in 
I_N$, $N\in \NN$, and $|{\cal I} \lambda | \leq 1$,
\be
| G_0^\lambda (z, z') | \leq C^N \kappa N^N | \Gamma (- {\cal R}\lambda )| 
(1 + \ln (2 \kappa | z - z' |^2)
\ee
for $2 \kappa  | z - z' |^2 < 1$ and
\be
| G_0^\lambda (z, z') | \leq C^N \kappa N^{2N} | \Gamma(- {\cal R} \lambda) | e^{- {\kappa \over 
2} | z - z' |^2}
\ee
for $2 \kappa | z - z' |^2 \geq 1$.
\par \noindent
Therefore if $\lambda \in I_N$, $N\in \NN$, $| \epsilon | <1$,
$\kappa > 2$ and $ n, n' \in \ZZ [i]$ with $n\neq n'$,
\bea
| (G_0^\lambda G_0^{\lambda + i \epsilon}) (n, n') | 
&\leq & C^{2N}\kappa^2 | \Gamma (-\lambda)|^2 N^{3N}\non \\
& & \ \ \ \ \ \  \times 
\Bigl \{\int_{2 \kappa | z - n |^2 < 1} dz (1 + \ln (2\kappa | z - n |^2) 
e^{- {\kappa \over 2} | z - n' |^2} \non \\
& + & \ \ \ \ \ \ \ 
\int_{2 \kappa | z - n' |^2 < 1} dz 
(1 + \ln (2 \kappa | z - n' |^2) e^{- {\kappa \over 2} | z - n |^2}  \non \\
& + & N^N \int dz e^{- {\kappa \over 2} | z - n|^2} e^{- {\kappa \over 2} 
| z - n' |^2} \Bigr \}  \non \\
& \leq & 2\pi C^{2N}  \kappa | \Gamma (- \lambda) |^2 N^{3N} 
e^{-{\kappa \over 8} | n - n' |^2}\!\!  
\left\{ \int^1_0 \!\!(1 + \ln r^2) rdr + N^N 
e^{- {\kappa \over 8}}\right\}  \non \\
& \leq & 2 C^{2N} \kappa^2 | \Gamma (- \lambda) |^2 N^{4N} 
e^{- {\kappa \over 8} | n - n' |^2}  \non \\
& \leq & e^{- {\kappa \over {32}}} C^{2N} | \Gamma (- \lambda) |^2 N^{4N} 
e^{- {1 \over 8} | n - n' |^2} 
\eea
if $\kappa$ is large enough.
Therefore
\be
|| Y || \leq \epsilon e^{- {\kappa \over {32}}} C^{2N} | \Gamma (- 
\lambda) |^2 || T ||
\ee
where $T$ is the operator with matrix 
$\la n | T | n' \ra = e^{- {1/8} | n - n' |^2}$.
Now take $\lambda \in (N - 1, N - \half]$ and $| \epsilon | < \lambda - N 
+ 1$. 
In this interval
\be
| \Gamma (- \lambda) | \leq  {{a_N} \over {(\lambda - N + 1)}} 
\ee
On the other hand by \cite{AS} 6.3.16
\be
{\cal I} \psi (- (\lambda + i \epsilon)) = - \epsilon \sum^\infty_{k = 0} {1 \over 
{(\lambda - k)^2 + \epsilon^2}} 
\ee
Therefore
\bea
| {\cal I}\psi (- (\lambda + i \epsilon))|  
& = & | \epsilon | \sum^\infty_{k = 0} {1 \over {(\lambda - k)^2 + \epsilon^2}}   >{{| \epsilon |} \over {(\lambda - N + 1)^2 + \epsilon^2}}  \non \\
& > & {{| \epsilon |} \over {2(\lambda - N + 1)^2}}.  
\eea
\par \noindent
Thus
\be
|| X^{-1} Y || \leq  || X^{-1} || || Y || 
\leq {\pi \over {4 \kappa}} a^2_N C^{2N} e^{- {\kappa \over  {32}}} || T || < 1 
\ee
if $\kappa$ is sufficiently large. Thus $M^{\lambda + i \epsilon }$              
is invertible. We can use the same argument if $\lambda \in [N - \half, N)$ and 
$ | \epsilon | < N - \lambda $.
\par
Using the bounds in Proposition 6.1, a similar calculation to the above gives for $\lambda \in I_0$,
\be
|| Y || \leq \epsilon e^{- {\kappa \over {32}}} C^{2N} | \lambda |^{-2} || T ||.
\ee 
Then using the inequality 
\be
| {\cal I}\psi (- (\lambda + i \epsilon))|  
>{{| \epsilon |} \over {|\lambda|^2 + \epsilon^2}},   
\ee
we can show that $M^{\lambda + i \epsilon }$ is invertible if 
$ | \epsilon | < |\lambda| $.
\qed
\par \noindent
{\bf Lemma 5.6} \ {\sl For $n \in \ZZ [i]$ and $\lambda \in J_N \setminus I_N$, let 
$\phi^\lambda_n = c_{\lambda, n} U_\lambda^* \Gamma^\lambda | n \ra$ where 
$c_{\lambda, n} = || U_\lambda^* \Gamma^\lambda | n \ra ||^{-1}$ so that 
$|| \phi_n^\lambda || = 1$. Then if $[a,b]\subset I_N$, the set $\{\phi^\lambda_n: n \in \ZZ [i], \ \lambda \in (J_N \setminus I_N )\cap \QQ [i]\}$ is total in ${\bf E} ([a,b]) {\cal H}$. }              
\par \noindent
{\bf Proof:}\ 
For $n \in \ZZ [i]$ and $\lambda \in J_N$ let
${\tilde \phi}_n^\lambda = U^*_\lambda | n \ra $.
Then if $\lambda \in J_N \setminus I_N$,
\be
{\tilde \phi}^\lambda_n = \sum_{n' \in \ZZ [i]} c^{-1}_{\lambda, n'} \la 
n' | M^\lambda | n \ra \phi^\lambda_{n'}.
\ee
Also if $\lambda \to \lambda'$ then $ {\tilde \phi}^\lambda_n \to 
{\tilde \phi}_n^{\lambda'}$.
Therefore it is sufficient to prove that the set $\{{\tilde \phi}^\lambda_n: n \in \ZZ 
[i], \lambda \in I_n\}$ is total. We do this by showing that the orthogonal 
complement of this set is in the orthogonal complement of ${\bf E} ([a,b]) 
{\cal H}$.
\par 
Let $f  \in {\cal H}$ and suppose that $({\tilde \phi}^\lambda_n, f) = 0 $
for all $n \in \ZZ [i]$ and all $\lambda \in  [a,b]$. Then since $(G^\lambda_0 
f) (n) = ({\tilde \phi}^\lambda_n, f)$, $G^\lambda f = G^\lambda_0 f$. 
Therefore ${\bf E} ([a,b]) G^\lambda f = {\bf E} ([a,b]) G^\lambda_0 f$ and 
thus 
\be
\sup_{\lambda \in [a,b]} || {\bf E} ([a,b]) G^\lambda f || \leq \sup_{\lambda 
\in [a,b]} || G^\lambda_0 || || f || < \infty.
\ee
Let $ \mu_1 (A) = (f, {\bf E} ([a,b] \cap A) f)$.
Then
\be
|| {\bf E} ([a,b])G^\lambda f ||^2 = \int_{[a,b]}  {{\mu_1 (d \lambda')} 
\over {| \lambda - \lambda' |^2}} .
\ee
Let $x_i  = a + (b - a)i/M$, $i = 0,\ldots, M $ and
$\lambda_i  = \half (x_i + x_{i+1})$.
Then
\be
\int_{[a,b]} {{\mu_1 (d \lambda')} \over {| \lambda' - \lambda_i |^2}} 
\geq \int_{[x_i, x_{i + 1}]} {{\mu_1 (d\lambda')} \over {| \lambda' - 
\lambda_i |^2}} \geq {{4M^2} \over {(b - a)^2}} \mu_1 ([x_i, x_{i + 1}])
\ee
Therefore
\be
\sup_{\lambda \in [a,b]} \int_{[a,b]} {{\mu_1(d \lambda')} \over {| 
\lambda' - \lambda |^2}} \geq {{4M^2} \over {(b-a)}} \mu_1([x_i, 
x_{i+1}])
\ee
for all $i$, and so
\be
\sup_{\lambda \in [a,b]} \int_{[a,b]} {{\mu_1 (d \lambda')} \over {| \lambda -\lambda'|^2}} 
\geq {{4M^2} \over {(b-a)^2}} {1 \over M} \sum^M_{i=0} 
\mu_1 ([x_i, x_{i+1}] 
\geq {{4M} \over {(b-a)^2}} \mu_1 ([a,b]).
\ee
Since $M$ is arbitrary
$\sup_{\lambda \in [a,b]} || {\bf E} ([a,b]) G^\lambda f || = \infty$
unless $\mu_1 ([a,b]) = 0$. 
But
$\mu_1 ([a,b]) = || {\bf E} ([a,b]) f ||^2$.
\qed\par
Let ${\cal F}$ be the  $\sigma$-algebra generated by
$\{\omega_{n'}: n'\in \ZZ[i]\}$ and
let ${\cal F}_n$ be the sub $\sigma$-algebra generated by
$\{\omega_{n'}: n'\neq n\}$.
Let ${\cal B}_N$ be the Borel sets of $I_N$.
\par\noindent
{\bf Lemma 5.7}\ {\sl Let $ B\mapsto {\bf E}(B) $ be the spectral measure of $H$ and
\break\hfill
$ A\in \cap_{n\in \ZZ[i]}({\cal F}_n\otimes{\cal B}_N)$.
If for a.e $\lambda \in I_N $ with respect to Lebesgue
measure $\EE \left \{ {\bf 1}_A(\ \cdot\ ,\lambda)\right \}=0$,  then
$\EE \left \{{\bf E}(\{\lambda:\ ( \ \cdot\ ,\lambda)\in A\})\right \}=0$.}
\par\noindent
{\bf Proof:}\ If for a.e $\lambda$ with respect to Lebesgue measure 
$\EE \left \{ {\bf 1}_A(\ \cdot\ ,\lambda)\right \}=0$, then by Fubini's Theorem
$\EE \left \{ \int_{I_N} d\lambda {\bf 1}_A(\ \cdot\ ,\lambda)\right \}=0$.
\par 
Let $\Lambda$ be a bounded subset of $\ZZ[i]$ and let $H_\Lambda$ be defined 
in the same way as $H$ with ${\cal M}$ replaced by ${\cal M}_\Lambda = l^2(\Lambda)$. By 
the same argument as in Proposition 4.1
$\lambda \notin \NN_0$ is an eigenvalue of $H_\Lambda$ if and only if there 
exists $v \in {\cal M}_\Lambda $ such that $M_\Lambda^\lambda 
v = 0$, where $M_\Lambda^\lambda$ is the restriction of $M^\lambda $ to 
$ {\cal M}_\Lambda $. Then the corresponding eigenfunction is $U^*_\lambda v $. Since in 
the interval  $I_N$, $\psi$ is 
bijective  it is clear that there are $| \Lambda |$  eigenvalues in $I_N$.
\par 
Let $\lambda_1, \ldots , \lambda_{| \Lambda |}$ be the 
eigenvalues in $I_N$, say, and let 
$v_1, \ldots, v_{| \Lambda |}$ be 
the corresponding vectors such that $M_\Lambda^{\lambda_k} v_k = 0$. Let $u_n = 
1/\omega_n$. Then for $n \in \Lambda $ we get 
\be
{{d \lambda_k} \over {d u_n}} = -{{| \la v_k | n \ra |^2} \over 
{|| U^*_{\lambda_k} v_k ||}} 
\label{e.8}
\ee
If $M_\Lambda^{\lambda_k} v_k = 0$ and $\la v_k | n \ra = 0$ for a particular 
value of $u_n$ then $M_\Lambda^{\lambda_k} v_k = 0$ for all values of $u_n$. 
We shall see later that we can ignore these eigenvalues.
\par 
We see from equation (\ref{e.8}) that each $\lambda_k$ is a  monotonic decreasing 
function of $u_n$. Moreover as $u_n \to \pm \infty$, the $\lambda_k$'s 
become identical, except the value of $\lambda_k$ corresponding to the $v_k$ which 
tends to $| n \ra$ and this latter value of $\lambda_k$ decreases from 
$N$ to $N-1$ (respectively $-\infty $ if $N=0$) as $u_n$ increases from 
$-\infty$ to $+\infty$. Therefore 
\be
\sum_k \int_{-\infty}^\infty f(\lambda_k) {{d \lambda_k} \over {d u_n}} du_n 
= -\int_{I_N} f(\lambda) d \lambda.
\label{e.9}
\ee
Let $\psi_k = {{U^*_{\lambda_k} v_k} \over {|| U^*_{\lambda_k} v_k 
||}}$ so that $H_\lambda \psi_k = \lambda_k \psi_k$. 
Let $\lambda \in J_N\setminus I_N$ and 
$n \in \ZZ [i]$. 
For $B\subset I_N$ let 
$\mu^{n,\lambda}_\Lambda (B) = (\phi^\lambda_n, {\bf E}_\Lambda (B) \phi^\lambda_n)$ 
where ${\bf E}_\Lambda$  is the spectral measure of 
$H_\Lambda$. Then for $\Lambda$ sufficiently large
\bea 
\mu^{n,\lambda}_\Lambda (B)  
& = &\sum_{\lambda_k \in B} | (\phi^\lambda_n, \psi_k) |^2 \non \\
& = &|| \Gamma^\lambda | n \ra ||^{-2} \sum_{\lambda_k \in B}
{1 \over {| \lambda - \lambda_k |^2}}
{{| \la n | v_k \ra |^2} \over 
{|| U^*_{\lambda_k} v_k ||^2}} \non \\
& = & - || \Gamma^\lambda        | n \ra ||^{-2} \sum_{\lambda_k \in B}
{1 \over {| \lambda - \lambda_k|^2}} {{d \lambda_k} \over {d u_n}}.
\label{e.10}
\eea
Note that if $\la n | v_k \ra=0$ then the corresponding term in (\ref{e.10})is absent. Also if $\lambda_k$ is degenerate, we can choose the corresponding orthogonal set of eigenvectors so that only one satisfies $\la n | v_k \ra
\neq 0 $. Therefore there is only one term corresponding to such $\lambda_k$ in the sum (\ref{e.10}). From (\ref{e.10}) and (\ref{e.9}) we get
\be
\int_{-\infty}^\infty du_n \mu^{n,\lambda}_\Lambda (B) 
=  || \Gamma^\lambda | n \ra ||^{-2} \int_B 
{{d \lambda'} \over {| \lambda - \lambda' |^2}},
\ee
and thus
\be
\int_{-\infty}^\infty du_n \rho(u_n) \mu^{n,\lambda}_\Lambda (B) 
\leq  || \Gamma^\lambda | n \ra ||^{-2} || \rho ||_\infty \int_B 
{{d \lambda'} \over {| \lambda - \lambda' |^2}}.
\label{e.11}
\ee
If $\mu^{n,\lambda}(B) = (\phi^\lambda_n, {\bf E}(B) \phi^\lambda_n)$ then by the weak convergence of $\mu^{n,\lambda}_\Lambda $ to $\mu^{n,\lambda}$ we have the bound (\ref{e.11}) for $\mu^{n,\lambda}$. By Kotani's argument \cite{K}, we have that
\hfill\break
$\EE \left \{ \int_{I_N} d\mu^{n,\lambda}(\lambda') {\bf 1}_A(\ \cdot\ ,\lambda')\right \}=0$. By Lemmas 5.4 and 5.6 we get that 
\hfill\break
$\EE \left \{{\bf E}(\{\lambda:\ ( \ \cdot\ ,\lambda)\in A\})\right \}=0$.}
\qed
\par
By combining Lemmas 5.3 and 5.7 we obtain our final theorem. 
\par\noindent 
{\bf Theorem 5.8}\ {\sl For each $N\in\NN$ there exists $\kappa_0 > 0$ 
such that for $\kappa >\kappa_0$, with probability one, 
$\sigma_{{\rm cont}}(H)\cap (-\infty, N)=\emptyset $,
and if $\lambda \in \sigma(H)\cap (-\infty,N)\setminus \NN_0$, 
is an eigenvalue of $H$ and the corresponding eigenfunction is 
$\phi_\lambda $, then for any compact subset $B$ of $\CC$, 
$\int_B|\phi_{\lambda}(z-z')|^2dz'$ decays exponentially in 
$z$ with exponential length less than or equal to $ 2/\kappa$.}
\vskip.25cm \noindent
{\bf Acknowledgements}
\par 
This work was supported by the Forbairt (Ireland) International Collaboration Programme 1997.
J.V.P. and T.C.D. would like to thank 
the Institut de physique th\'eorique of the Ecole Polytechnique
F\'ed\'erale de Lausanne for their hospitality and financial support.
T.C.D. and N.M. would like to thank University College Dublin for their hospitality. 
%
\section{ Appendix A.\ Bounds for the Green's Function}
In this appendix we shall obtain bounds on the Green's function 
$G^\lambda_0(z,z')$. Our basic tools are the the integral 
representation (\cite{AS} 13.2.5)
\be
\Gamma (a) U(a, b, \rho) = \int^\infty_0 dt e^{- \rho t } 
t^{a-1} (1 + t)^{b-a-1},  \label{f.1}
\ee
which is valid for ${\cal R}a > 0$ and $ \rho > 0$ and
the recurrence relation (\cite{AS} 13.4.18)
\be
U(a,b,\rho) = \rho U(a+1, b+1, \rho) - (b-a-1) U(a+1, b,\rho). 
\label{f.2}
\ee 
We first obtain bounds for $| G_0^\lambda (z,z') | $ when 
${\cal R}\lambda <0$.
\par\noindent
{\bf Proposition 6.1} \ {\sl There exists a constant $C<\infty$,
such that for ${\cal R}\lambda \in I_0$, 
\be
|  G^\lambda_0(z,z') | \leq
C \kappa e^{- \kappa | z - z' |^2}
\left \{ {1\over{|{\cal R }\lambda|}}+ 
e^{-\sqrt{2\kappa |{\cal R}\lambda|} |z-z'|} 
\(1+|\ln \(2\kappa | z - z' |^2\)|\)\right \},
\ee
if $2\kappa | z - z' |^2 \leq 1$, and
\be
|  G^\lambda_0(z,z') | \leq
{{C \kappa}\over{|{\cal R }\lambda |}}
e^{- \kappa | z - z' |^2},
\ee 
if $2\kappa | z - z' |^2\ > 1$. 
}
\par \noindent
{\bf Proof:}\ 
Let $\lambda =  x + iy $ with $ x < 0$. Then from (\ref{f.1}) we get
\bea|\Gamma (-\lambda) U(-\lambda, 1, \rho)| 
&\leq &\int^{\infty}_0 dt e^{- \rho t} t^{|x|-1} (1 + t)^{-|x|} \non \\
& = &\int^1_0 dt e^{- \rho t} t^{|x|-1}(1+t)^{-|x| }
+  \int^{\infty}_1 dt e^{- \rho t} t^{|x|-1}(1 + t)^{-|x|} \non \\
& \leq &\int^1_0 dt\ t^{|x|-1}  + \int^{\infty}_1 dt 
{{e^{-\rho t}} \over t}
\left({t \over {1+t}}\right)^{|x|} \non \\
&\leq &  {1 \over |x|}  + \int^{\infty}_1 dt {{e^{- (\rho t +{{|x|}\over {2 t}})}} \over t}.
\eea
If $\rho \leq 1$, we have 
\bea
|\Gamma (-\lambda) U(-\lambda, 1, \rho)| 
&\leq &  {1 \over |x|}  + \int^{\infty}_1 dt {{ e^{- {\half\rho t}}
e^{- \half(\rho t+{{|x|}\over { t}})}} \over t}\non \\
&\leq &  {1 \over |x|}  + e^{- (\rho |x|)^\half } 
\int^{\infty}_1 dt {{ e^{- {\half\rho t}}}\over t} \non \\
& = & {1 \over |x|}  + e^{- (\rho|x|)^\half }\int^{\infty}_{\half\rho} dt 
{{e^{-t}} \over t} \non \\
& \leq & {1 \over |x|} + e^{- (\rho|x|)^\half }\int^{1}_{\half\rho} {{dt} \over t}  
+ e^{- (\rho|x|)^\half }\int^{\infty}_{1} dt {{e^{-t}} \over t} \non \\
& \leq & {1 \over |x|} - e^{-(\rho |x|)^\half }\ln (\rho/2) + e^{- (\rho|x|)^\half }\int^{\infty}_1 dt e^{-t} \non \\
& = & {1 \over |x|}  + e^{- (\rho |x|)^\half }| \ln (\rho/2) | + {{ e^{- (\rho|x|)^\half }} \over e}. \non 
\eea
Thus, 
\be
|\Gamma(-\lambda) U(-\lambda, 1, \rho) |
\leq C\left \{ {1 \over |x|}+\left (1 + | \ln \rho |\right ) e^{- (\rho |x|)^\half }\right \}.
\label{f.3}
\ee

If $\rho > 1$, we have 
\bea
|\Gamma (-\lambda) U(-\lambda, 1, \rho)| 
&\leq &  {1 \over |x|}  + \int^{\infty}_1 dt {{e^{- (t +{{|x|}\over {2 t}})}} \over t}\non \\
&\leq &  {1 \over |x|}  + \int^{\infty}_1 dt {{ e^{- {\half t}}
e^{- \half(t+{{|x|}\over { t}})}} \over t}\non \\
&\leq &  {1 \over |x|}  + e^{- |x|^\half }
\int^{\infty}_1 dt  e^{- {\half t} }  \non \\
&= & {1 \over |x|}  + 2 e^{- |x|^\half } \non 
\eea
Therefore, 
\be
|\Gamma(-\lambda) U(-\lambda, 1, \rho) |
\leq {C \over |x|}.
\label{f.4}
\ee
Inserting the inequalities (\ref{f.3}) and (\ref{f.4}) into (\ref{b.0}) we 
get Proposition 6.1. 
\qed
\par
Now we shall obtain bounds for ${\cal R}\lambda >0$.
\par\noindent
{\bf Proposition 6.2} \ {\sl There exists a constant $C<\infty$,
such that for ${\cal R}\lambda \in I_N$, $N\in \NN$,$|{\cal I}\lambda |\leq 1$,
\be
| G^\lambda_0 (z,z') | 
\leq \kappa C^N N^N  | \Gamma (-{\cal R}\lambda)| 
(1 + | \ln (2\kappa | z  - z' | ^2)|)
e^{- \kappa | z - z' |^2},
\ee
if $2\kappa | z - z' |^2 \leq 1$, and
\be
| G^\lambda_0(z,z') | 
\leq(C\kappa)^{N+1} N^N  | \Gamma (-{\cal R}\lambda)|
| z - z' |^{2N} e^{- \kappa | z - z' |^2},
\ee 
if $2\kappa | z - z' |^2\ > 1$. 
}
\par
Let $\lambda =x+iy$. We shall prove that if $N-1 < x < N $, $N\in \NN_0$, 
$b \in \NN$ and $ \rho > 1$, then
\be
| U(-\lambda,b,\rho) | 
\leq  2^{b + N - 1}\rho^x 
(b + N+|y|)^N \left | {{\Gamma(-x)}\over {\Gamma(-\lambda)}}\right |
+ e^{-(\rho-2)} (\rho+|y|+1)^N  
{{(b+N)!} \over {|\Gamma (N-\lambda)|}}.
\label{f.6}  
\ee 
We shall do this by induction on $N$. We first prove (\ref{f.6}) for $N=0$ 
by using (\ref{f.1}) which gives 
\bea
|\Gamma (-\lambda) U(-\lambda, b, \rho)| 
& \leq &\int^1_0\!\!\!\! 
dt e^{-\rho t} t^{-(x+1)} (1+t)^{b+x-1} + 
\int^\infty_1\!\!\!\!\!\! 
dt e^{-\rho t} t^{-(x+1)} (1 +t)^{b+x-1} \non \\
& = &I_1 + I_2. 
\non
\eea
We now take $\rho > 1$, $-1<x<0$ and $b\geq 1$.
For $I_1$, since $t<1$, we get
\bea
I_1 
& \leq &2^{b-1}\int^1_0 dt e^{-\rho t} t^{-(x+1)}
= 2^{b-1}\rho^x \int^\rho_0 dt e^{-t} t^{-(x+1)}  \non \\
& \leq & 2^{b-1}\rho^x \int^\infty_0 dt e^{-t} t^{-(x+1)} 
= 2^{b-1}\rho^x \Gamma (-x). \non 
\eea
On the other hand, using $t > 1$, we get
\bea
I_2 
& = &\int^\infty_1 \!\!\!\!\!\!
dt e^{-(\rho-1)t} 
e^{-t} t^{-(x+1)} (1+t)^{b+x-1} \leq e^{-(\rho-1)} 
\int^\infty_1 \!
\!\!\!\!\!
dt e^{-t} t^{-(x+1)} (1 +t)^{b+x-1} \non \\
& \leq & 
e^{-(\rho-1)} \int^\infty_1 dt e^{-t} (1 + t)^{b-1}
. \non 
\eea
Therefore
\be
I_2 \leq e^{- (\rho-1)} \int^{\infty}_2\!\!  ds e^{- s+1} s^{b-1} 
\leq e^{-(\rho-2)} \int^\infty_0 \!\!ds e^{-s} s^{b-1} 
= e^{-(\rho-2)} \Gamma (b)\leq e^{-(\rho-2)}b!. 
\ee
Thus we have 
\be
|\Gamma (-\lambda) U(-\lambda, b, \rho) |
\leq 2^{b-1}\rho^x \Gamma(-x) + e^{-(\rho-2)} b!
\label{f.7}
\ee 
for $ \rho > 1$ and $ -1 < x < 0 $, or
\be
| U(-\lambda,b,\rho) | 
\leq 2^{b-1}\rho^x \left | {{\Gamma(-x)}\over {\Gamma(-\lambda)}}\right |
+ e^{-(\rho-2)} {{b!} \over {|\Gamma(-\lambda)|}}. 
\label{f.8}
\ee                           
Suppose that (\ref{f.6}) is true for 
$ N-1 < x <  N$. Then by using the recurrence relation (\ref{f.2}),
we get for $ N < x < N+1$
\bea
| U(\!\!\!\!\!&-&\!\!\!\!\! \lambda ,b,\rho)|\non \\  
&\leq &
\rho \Bigl\{2^{b+N}\rho^{(x-1)} (b + N + |y|+1)^N
\left | {{\Gamma(-x+1)}\over {\Gamma(-\lambda +1)}}\right | \non \\
& & \ \ \
+ e^{-(\rho-2)} (\rho+|y|+1)^N {{(b+N+1)!} 
\over {|\Gamma(N -\lambda+ 1)|}}\Bigr\}\non \\
& & \ \ \ + |b +\lambda-1| \Bigl \{2^{b + N -1}\rho^{(x-1)}(b + N+|y|)^N 
\left | {{\Gamma(-x+1)}\over {\Gamma(-\lambda +1)}}\right | \non \\
& & \ \ \ \ \  
+ e^{-(\rho-2)}
(\rho+1+|y|)^N {{(b+N)!} \over {|\Gamma(N-\lambda+ 1)|}}\Bigr \}.\non 
\eea
The identity $\Gamma(x+1)=x\Gamma(x) $ gives
$$
\left |{{\Gamma(-x+1)}\over{\Gamma(-\lambda+1)}}\right| =
\left |{{x}\over{\lambda}}\right|
\left |{{\Gamma(-x)}\over{\Gamma(-\lambda)}}\right|
\leq \left |{{\Gamma(-x)}\over {\Gamma(-\lambda)}}\right|.
$$ 
Therefore
\bea
| U(-\lambda,b,\rho)|\!\!  
&\leq &\!\! 2^{b+N}\rho^x \left\{(b+ N + |y|+1)^N + {{|b+\lambda -1|} 
\over {2\rho}} (b + N+|y|)^N\right\}
\left | {{\Gamma(-x)}\over {\Gamma(-\lambda)}}\right | \non \\
& & \ \ \ \ \ \ \ \ \ \ + e^{-(\rho-2)} (\rho+|y|+1)^N {{(b+N+1)!}
\over {|\Gamma(N-\lambda+1)|}}
\left \{ \rho +{{|b+\lambda -1|}\over {b+N+1}}\right\} \non \\
& \leq & \!\! 2^{b + N}(b + N + |y| +1)^N \rho^x\left\{1 + {{|b+\lambda-1|} \over 
{2\rho}} \right\}
\left | {{\Gamma(-x)}\over {\Gamma(-\lambda)}}\right | \non \\
& & \ \ \ \ \ \ \ \ \ \ + e^{-(\rho-2)} (\rho+|y|+1)^{N+1} {{(b+N+1)!} 
\over {\Gamma(N-\lambda+1)}}. \non 
\eea
Therefore since
$2 \leq b +  N + |y|+1$ and
$|b+\lambda-1 |\leq b +  N + |y|$
we get the required bound 
$$
| U (-\lambda, b, \rho) | 
\leq 2^{b+ N}\rho^x (b + N + |y|+1)^{N+1} 
\left | {{\Gamma(-x)}\over {\Gamma(-\lambda)}}\right | 
+ e^{-(\rho-2)} (\rho +1)^{N+1} {{(b+N+1)!} \over {\Gamma(N-\lambda+1)}}. 
$$    
This gives the following bound on the Green's function 
for $2 \kappa | z - z' |^2 \geq 1$ and
$N - 1 < x < N $ and $|y|\leq 1$
\bea
|\!\!\!\!\!\!\! & & \!\!\!\!\!\!\! 
G^\lambda_0(z,z') | \leq {{2 \kappa} \over \pi}  
e^{- \kappa | z - z' |^2} | \Gamma (- \lambda) |  \non \\
&\times &
\Bigl \{(2 \kappa)^x (2(2 + N))^N | z - z' |^{2x} 
\left | {{\Gamma(-x)}\over {\Gamma(-\lambda)}}\right | \non \\
& & \ \ \ \ \ 
+ e^2(2 + 2 \kappa | z  - z' |^2)^N 
{{(N+ 1)!} \over {|\Gamma (N - \lambda)|}} 
e^{- 2 \kappa | z - z' |^2}\Bigr \}.  
\label{f.9}
\eea
From (\ref{f.9}) we get for $N - 1 < x < N $ with $N\in \NN_0$
\be
| G^\lambda_0(z,z') | 
\leq(C\kappa)^{N+1} N^N  | \Gamma (-x)|
| z - z' |^{2N} e^{- \kappa | z - z' |^2}   
\ee
since $| \Gamma(-\lambda)|\leq | \Gamma(-x)|$ and $ \Gamma (N - \lambda)$
is bounded below. 
\par
We shall prove, again by induction, that for $N-1 < x <N$, $N\in \NN_0$, 
$b\in \NN $ and $\rho\leq 1$
\be
|U(-\lambda,b,\rho)| \leq 2^{N+4} (b+N)! 
\left | {{\Gamma (-x)}\over {\Gamma (-\lambda)}}\right |
{{(1 + | \ln \rho |)} \over {\rho^{b-1}}}.
\label{f.13}
\ee
We first prove (\ref{f.13}) for $N=0$.
From (\ref{f.4}) we have for $\rho \leq 1$, and $x < 0$,
\be
|\Gamma(-\lambda) U(-\lambda, 1, \rho)| 
\leq {1 \over {|x|}} + {1 \over e} + | \ln \rho |.
\ee
Thus
\be
|U(-\lambda, 1 ,\rho)| 
\leq 
\left | {{\Gamma (-x)}\over {\Gamma (-\lambda)}}\right |
\left \{{1 \over {{|x|} \Gamma (-x)}} 
+ {1 \over {e \Gamma (-x)}} 
+ {1 \over {\Gamma (-x)}} | \ln \rho |\right \}.
\ee
Since $ -1 < x < 0$, $\Gamma (-x) > 1$
and $|x| \Gamma (-x) = \Gamma (-x+1)> (e-1)/e $, this gives
\bea
\label{f.14}
|U(-\lambda,1, \rho)| 
&\leq &
\left | {{\Gamma (-x)}\over {\Gamma (-\lambda)}}\right |
\left \{ {{e^2+e-1}\over {e^2-e}} + | \ln \rho | \right \}\\ \non
&\leq &
(2 + | \ln \rho |) \left | {{\Gamma (-x)}\over {\Gamma (-\lambda)}}\right|
\leq 2^4\left | {{\Gamma (-x)}\over {\Gamma (-\lambda)}}\right |
(1 + | \ln \rho |).
\eea
\par
Now we take $b \geq 2$.
\bea
|\Gamma (-\lambda) U(-\lambda,b,\rho) |
& \leq &\int^{\infty}_0 dt e^{-\rho t} t^{-(x+1)} (1 +t)^{b+x-1}\non \\
& = & {1 \over {\rho^{b-1}}} \int^{\infty}_0 dt e^{-t} t^{-(x+1)} 
(\rho +t)^{b+x-1}\non \\
& \leq &{1 \over {\rho^{b-1}}} \int^{\infty}_0 dt  e^{-t} 
t^{-(x+1)} (1 + t)^{b+x-1}\non 
\eea
since  $b +x- 1 \geq 0$.
Thus we have for $b \geq 2$, and $-1 < x < 0$
\be
|\Gamma (-\lambda) U(-\lambda,b,\rho)| 
\leq {1 \over {\rho^{b-1}}} \Gamma (-x) U(-x, b, 1).
\ee
By inserting the bound obtained from (\ref {f.7}) by letting $\rho$ tend to $1$,
\be
\Gamma (-x)U(-x, b, 1) \leq 2^{b-1}\Gamma (-x) + e b!,
\ee
into this inequality we get
\be
 |U(-\lambda,b,\rho)| 
\leq {1 \over {\rho^{b-1}}} \left\{2^{b-1} + {{e b!}\over{\Gamma (-x)}} \right\}\left | {{\Gamma (-x)}\over {\Gamma (-\lambda)}}\right |
\leq 2^4 b! 
\left | {{\Gamma (-x)}\over {\Gamma (-\lambda)}}\right |
{{(1 + | \ln \rho |)} \over {\rho^{b-1}}}. 
\ee
We can combine this with the inequality (\ref{f.14}) 
to get for $b\in \NN$, $-1 < \lambda < 0$ and $ \rho < 1$,
\be
|U(-\lambda,b,\rho)| 
\leq 2^4 b! 
\left | {{\Gamma (-x)}\over {\Gamma (-\lambda)}}\right |
{{(1 + | \ln \rho |)} \over {\rho^{b-1}}}
\ee
Using the recurrence relation (\ref{f.2}) and the induction hypothesis we get for
$N< \lambda < N+1$,
\bea
|U(-\lambda,b,\rho)| 
&\leq &\rho \left\{\leq 2^{N+4} (b+N+1)! 
\left | {{\Gamma (-x+1)}\over {\Gamma (-\lambda+1)}}\right |
{{(1 + | \ln \rho |)} \over {\rho^b}} \right\}\non \\
& & \ \ \ \  \ \ \ \ \ \  \ \ \ \ 
+ |b+\lambda -1|2^{N+4} (b+N)! 
\left | {{\Gamma (-x+1)}\over {\Gamma (-\lambda+1)}}\right |
{{(1 + | \ln \rho |)} \over {\rho^{b-1}}}\non \\
&\leq & 2^{N+4} (b+N)! 
\left | {{\Gamma (-x+1)}\over {\Gamma (-\lambda+1)}}\right |
{{(1 + | \ln \rho |)} \over {\rho^{b-1}}}
(b+N+1+|b+\lambda -1|)\non \\
&\leq & 2^{N+5} (b+N+1)! 
\left | {{\Gamma (-x)}\over {\Gamma (-\lambda)}}\right |
{{(1 + | \ln \rho |)} \over {\rho^{b-1}}}
\eea
In particular with $b = 1$, (\ref{f.13}) gives for
$N-1 < x < N $ and $\rho \leq 1$
\be
|U(-\lambda, 1, \rho) |
\leq 2^{N+4} (N+1)! 
\left | {{\Gamma (-x)}\over {\Gamma (-\lambda)}}\right |
(1 + | \ln \rho |).
\ee
This gives us the required bound on the Green's function 
for $ N - 1 < x < N$, and $2 \kappa | z  - z' |^2 \leq 1$
\bea
| G^\lambda_0 (z,z') | 
& \leq & {{2 \kappa} \over \pi} |\Gamma (- x)| 2^{N+4} (N+1)! 
(1 +  | \ln (2\kappa| z  - z' |^2) |)
e^{- \kappa | z - z' |^2}\non \\
& \leq & \kappa C^N N^N  | \Gamma (- x)| 
(1 + | \ln (2\kappa | z  - z' | ^2)|)
e^{- \kappa | z - z' |^2}.
\eea
\par
%
\section{Appendix B. Regularity of $\mu$}
{\bf Definition} {\sl A probability measure $\mu$ on $\RR$ is said to be $\tau$-regular,
with $\tau\in (0,1]$, if there exists $\nu >0$ and $C<\infty$ such that 
\be
\mu([x - \delta, \ x + \delta]) \leq C \delta^\tau \mu([x - \nu, x+\nu])
\ee
for all $x \in \RR$ and $0 < \delta < 1$.}
\par\noindent
Note that it is equivalent to requiring that there exists 
$\nu >0$ and $C<\infty$ such that 
\be
\mu([x - \delta, \ x + \delta]) \leq C \delta^\tau \mu([x - \nu, x+\nu])
\label{G.2}
\ee
for all $x \in \RR$ and $0 < \delta < \nu$.  We shall prove this with 
$\tau=1$.
Recall that the
probability measure $\mu_0$ has support is an interval 
$[-a,a]$ with $a <\infty $. $\mu_0$ is symmetric about the origin and that its density $\rho_0$ is differentiable on $(-a,a)$ and 
satisfies the following condition
\be
A \equiv \sup_{\zeta\in (0,a)} \frac{\rho_0'(\zeta)} {\rho_0(\zeta)}< \infty.
\ee
If $B\subset \RR$, $\mu(B)=\mu_0(\{\omega\ :\ 1/\omega \in B\})$ and the density of $\mu$, $\rho$,
is given by
\be
\rho (x)=\frac{\rho_0(1/x)}{x^2}. 
\ee 
Since in our case $\mu $ is symmetric about the origin, 
it is sufficient to prove (\ref{G.2}) for $x\geq 0$. 
Also it is easy to see that the following condition is sufficient for (\ref{G.2}) with $\tau=1$.
\par\noindent
{\sl There exists $\nu >0$ and $C<0$ such that 
\be
\rho (x+t')\leq C\rho (x+t) 
\label{G.3}
\ee
for all $x\in R_{+}$, $-\nu \leq t' \leq t \leq \nu $.} 
\par\noindent
Then
\bea
\mu ([x-\delta ,x+\delta ]) & = & \frac \delta \nu \int_{-\nu }^\nu \rho
(x+\frac \delta \nu t)dt \non \\  
& \leq  & \frac{\delta (C+1)}\nu \int_0^\nu \rho (x+\frac \delta \nu t)dt \non \\
& \leq  & \frac{\delta (C+1)C}\nu \int_0^\nu \rho (x+t)dt \non \\  
& = & \frac{\delta (C+1)C}\nu \mu ([x,x+\nu ])
\eea
Let $b=1/a$. If $0\leq x\leq b-t'$, then $\rho (x+t')=0$. If $x>b-t'$, then 
\be
\ln\rho_0(1/(x+t'))-\ln\rho_0(1/(x+t))
=\frac {t-t'}{(x+t)(x+t')}
\frac{\rho'_0(\zeta)}{\rho_0(\zeta)}, 
\ee
where $\zeta\in (1/(x+t),1/(x+t'))$. Thus
\be
\ln\rho_0(1/(x+t'))-\ln\rho_0(1/(x+t))\leq \max(0,\frac {2A\nu}{b^2}).
\ee
Therefore, with $C'=\exp(\max(0,\frac {2A\nu}{b^2}))$,
\be
\rho_0(1/(x+t'))\leq C'\rho_0(1/(x+t)).
\ee
But
\be
\left(\frac{x+t}{x+t'}\right)^2 
\leq\left(\frac{b+2\nu}b\right)^2. 
\ee
Thus the inequality (\ref{G.3}) is satisfied with 
$C= C'\left(\frac{b+2\nu}b\right)^2$. 
Finally note that 
\be
\int |x|^\epsilon \mu(dx)= \int |x|^{-\epsilon}\rho_0(x)dx< \infty
\ee
for all $\epsilon<1$ since $\rho_0$ is continuous at the origin. 
%

\end{document}